\documentclass[preprint,aps,prd,superscriptaddress,longbibliography]{revtex4-1}
\usepackage{amsthm,amsmath,amsfonts,amssymb,verbatim,color}
\usepackage{graphicx}
\usepackage{bm}
\usepackage{epsfig,slashed}
\usepackage{amssymb}
\usepackage{amsfonts}
\usepackage{tikz}
\usepackage{xparse}
\usepackage{ifthen}
\usepackage{lipsum}
\usepackage[colorlinks=true,citecolor=blue,linkcolor=blue,urlcolor=blue]{hyperref}
\usepackage[caption=false]{subfig}
\usepackage[papersize={8.5in,11in}]{geometry}
\usepackage{nccmath}
\usepackage{cleveref}

\begin{document}

\newcommand{\tr}{\mathop{\mathrm{tr}}}
\newcommand{\bsigma}{\boldsymbol{\sigma}}
\newcommand{\re}{\mathop{\mathrm{Re}}}
\newcommand{\im}{\mathop{\mathrm{Im}}}
\renewcommand{\b}[1]{{\boldsymbol{#1}}}
\newcommand{\diag}{\mathrm{diag}}
\newcommand{\sign}{\mathrm{sign}}
\newcommand{\sgn}{\mathop{\mathrm{sgn}}}
\renewcommand{\c}[1]{\mathcal{#1}}
\renewcommand{\mod}{\mathop{\mathrm{mod}}}
\renewcommand{\geq}{\geqslant}
\renewcommand{\leq}{\leqslant}

\newcommand{\cl}{\mathrm{cl}}
\newcommand{\mb}{\bm}
\newcommand{\ua}{\uparrow}
\newcommand{\da}{\downarrow}
\newcommand{\ra}{\rightarrow}
\newcommand{\la}{\leftarrow}
\newcommand{\mc}{\mathcal}
\newcommand{\bs}{\boldsymbol}
\newcommand{\lra}{\leftrightarrow}
\newcommand{\nn}{\nonumber}
\newcommand{\half}{{\textstyle{\frac{1}{2}}}}
\newcommand{\mf}{\mathfrak}
\newcommand{\MF}{\text{MF}}
\newcommand{\IR}{\text{IR}}
\newcommand{\UV}{\text{UV}}
\newcommand{\sech}{\mathrm{sech}}

% Added by Hena

\newcommand{\abs}[1]{ \lvert #1 \rvert\ }

% end Added by Hena

\title{Random-mass disorder in the critical Gross-Neveu-Yukawa models}

\author{Hennadii Yerzhakov}
\affiliation{Department of Physics, University of Alberta, Edmonton, Alberta T6G 2E1, Canada}

\author{Joseph Maciejko}
\affiliation{Department of Physics, University of Alberta, Edmonton, Alberta T6G 2E1, Canada}
\affiliation{Theoretical Physics Institute (TPI), University of Alberta, Edmonton, Alberta T6G 2E1, Canada}

\date{\today\\
\vspace{0.4in}}

\begin{abstract}
An important yet largely unsolved problem in the statistical mechanics of disordered quantum systems is to understand how quenched disorder affects quantum phase transitions in systems of itinerant fermions. In the clean limit, continuous quantum phase transitions of the symmetry-breaking type in Dirac materials such as graphene and the surfaces of topological insulators are described by relativistic (2+1)-dimensional quantum field theories of the Gross-Neveu-Yukawa (GNY) type. We study the universal critical properties of the chiral Ising, XY, and Heisenberg GNY models perturbed by quenched random-mass disorder, both uncorrelated or with long-range power-law correlations. Using the replica method combined with a controlled triple epsilon expansion below four dimensions, we find a variety of new finite-randomness critical and multicritical points with nonzero Yukawa coupling between low-energy Dirac fields and bosonic order parameter fluctuations, and compute their universal critical exponents. Analyzing bifurcations of the renormalization-group flow, we find instances of the fixed-point annihilation scenario---continuously tuned by the power-law exponent of long-range disorder correlations and associated with an exponentially large crossover length---as well as the transcritical bifurcation and the supercritical Hopf bifurcation. The latter is accompanied by the birth of a stable limit cycle on the critical hypersurface, which represents the first instance of fermionic quantum criticality with emergent discrete scale invariance.
\end{abstract}

\maketitle

\section{Introduction}

Understanding the effect of quenched disorder on continuous quantum phase transitions is a question of enduring interest~\cite{vojta2013,*vojta2019}, motivated by the ubiquitous presence of imperfections in the condensed matter systems that exhibit such transitions. In the clean limit, the divergence of the correlation length at criticality produces universal critical phenomena that are controlled by renormalization-group (RG) fixed points of a translationally invariant continuum quantum field theory. A given disorder configuration manifestly breaks translation symmetry even on long length scales, and thus produces behavior very different from that of a translationally invariant system. However, that symmetry is restored in physical properties upon averaging over such configurations. The cases of main interest, then, are those in which disorder qualitatively affects the long-distance physics even after disorder averaging. For quantum critical points (QCPs) described at long distances by a (2+1)D strongly interacting conformal field theory in the clean limit, such as many QCPs of interest in condensed matter physics~\cite{QPT}, determining the fate of the system in the infrared after disorder averaging is a problem fraught with technical difficulties. For example, thermodynamic properties are in principle determined by first computing the partition function of a strongly coupled quantum field theory with spatially random couplings, then averaging its logarithm over some chosen probability distribution. An approach better suited to determining the long-distance behavior of the system, our only concern here, is to investigate the RG flow of disorder-averaged observables. It was recently shown~\cite{narovlansky2018,aharony2018} that this is equivalent to studying the RG flow of an effective theory with disorder-induced translationally-invariant interactions, derived using the standard replica trick~\cite{emery1975}, despite the formally nonlocal nature of such theories and oft-invoked concerns about the validity of analytically continuing the number of replicas to zero.

\begin{figure}[t]
\includegraphics[width=0.9\columnwidth]{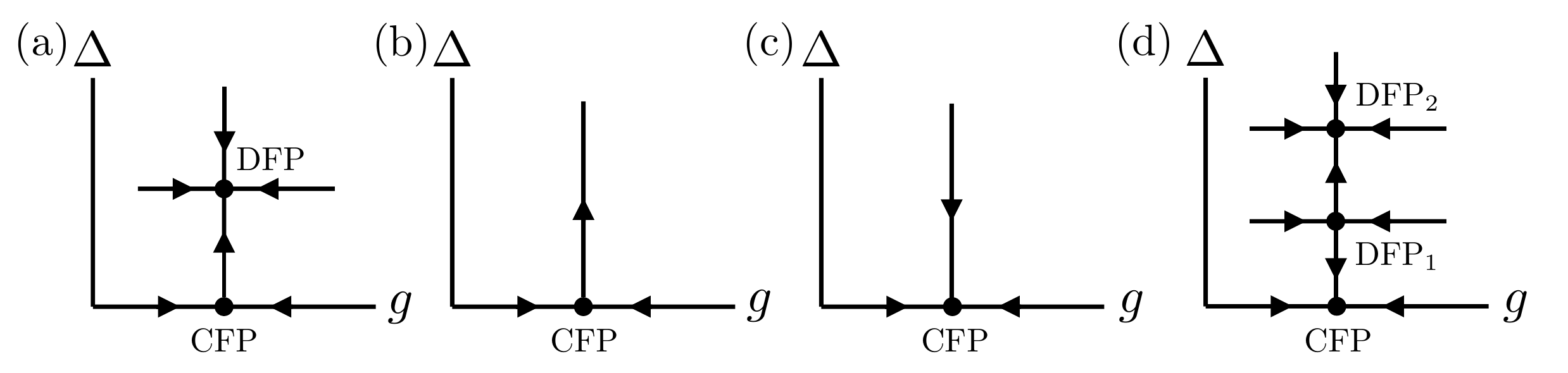}
\caption{Possible schematic RG flows involving disordered fixed points (DFP) on the critical hypersurface in the vicinity of a clean fixed point (CFP), corresponding to a conformal field theory with nonzero interaction strength $g$, perturbed by disorder $\Delta$.}
\label{fig:RG}
\end{figure}

A situation of particular interest is one in which disorder produces RG flows on the critical hypersurface that connect the clean fixed point (CFP) describing the transition in the absence of disorder to fixed points characterized by a nonzero value of the effective disorder coupling(s). Such disordered fixed points (DFPs) exhibit scaling behavior but, by contrast with CFPs, no self-averaging in the thermodynamic limit~\cite{aharony1996}. We will be exclusively concerned with random-mass disorder, also known as random-$T_c$ disorder in the context of classical (thermal) phase transitions, every configuration of which preserves those symmetries of the system that are broken spontaneously at the transition. (In $d=2$ spatial dimensions, the focus of this paper, random-field disorder---which violates those symmetries---precludes long-range order, and thus the possibility of a sharp transition~\cite{imry1975,aizenman1989,greenblatt2009,aizenman2012}.) The standard scenario is one in which random-mass disorder is a relevant perturbation at the CFP [Fig.~\ref{fig:RG}(a)], and drives a direct RG flow to a DFP. For short-range correlated disorder, this occurs when the correlation length exponent $\nu_\text{CFP}$ of the CFP obeys the Harris inequality $\nu_\text{CFP}<2/d$~\cite{harris1974}. Examples include the superfluid-Mott glass transition of bosons with particle-hole symmetry in $d=2$~\cite{vojta2016} and $d=3$~\cite{crewse2018}, described by the $O(2)$ vector model with random-mass disorder in (2+1)D ($\nu_\text{CFP}\approx 0.67<1$) and (3+1)D ($\nu_\text{CFP}=1/2<2/3$), respectively. The true correlation length exponent $\nu$ in the presence of disorder, i.e., its value at the DFP, obeys the Chayes inequality $\nu\geqslant 2/d$~\cite{chayes1986}; the dynamic critical exponent $z$ changes from its Lorentz-invariant value $z=1$ at the conformally invariant $O(2)$ Wilson-Fisher fixed point to some noninteger but equally universal (and finite) value $z>1$ at the DFP (see Table~\ref{tab:SFMG}). Finite-randomness DFPs in the $O(n)$ vector model are in principle accessible via perturbative RG analyses of the disorder-averaged effective field theory combined with (double) epsilon~\cite{Khmelnitskii1978,Dorogovtsev1980,Boyanovsky1982,boyanovsky1983,Lawrie1984} or $1/n$ expansions~\cite{goldman2020}. Infinite-randomness DFPs (for which $z=\infty$) are also possible, such as those describing the random-bond transverse-field Ising model at criticality in (1+1)D~\cite{fisher1992,*fisher1995} ($\nu_\text{CFP}=1<2$) and (2+1)D~\cite{motrunich2000} ($\nu_\text{CFP}\approx 0.63<1$). These, however, are not adequately captured by perturbative RG analyses of a disorder-averaged continuum field theory, given the runaway flow to infinite disorder [Fig.~\ref{fig:RG}(b)]. Rather, they can be quantitatively studied using strong-disorder real-space RG methods~\cite{ma1979,dasgupta1980,fisher1992,*fisher1995,motrunich2000} which, in spatial dimensions $d\geqslant 2$ at least, must be implemented numerically in microscopic lattice models.

If disorder is Harris-irrelevant, $\nu_\text{CFP}>2/d$, the standard lore is that disorder has no effect on the phase transition at long distances [Fig.~\ref{fig:RG}(c)]. However, as observed in Ref.~\cite{aharony2016}, an irrelevant perturbation with a finite coefficient can have nontrivial consequences on the RG flow finitely away from the CFP, just as formally irrelevant interactions at a stable noninteracting fixed point can eventually trigger a phase transition and produce a critical fixed point. The simplest possible RG flow leading to a DFP in the case of Harris-irrelevant disorder is illustrated in Fig.~\ref{fig:RG}(d), and was recently found in a double epsilon-expansion study of the random-mass chiral XY Gross-Neveu-Yukawa (GNY) model~\cite{Yerzhakov2018}, a fermionic analog of the $O(2)$ vector model that, absent disorder, describes the quantum phase transition between a Dirac semimetal and a gapped superconductor (Ref.~\cite{Roy2013,Zerf2016}, and also see Sec.~\ref{sec:model}). Below a separatrix line controlled by a disordered saddle-type fixed point (DFP$_1$), the transition is in the same universality class as the clean sytem, while above that separatrix line, the transition is governed by a disordered critical point (DFP$_2$). A similar RG flow is found in the classical 2D Ising model with binary ($\pm J$) random-bond disorder~\cite{picco2006}. For a weak concentration of antiferromagnetic bonds randomly distributed amidst ferromagnetic bonds, the paramagnetic-ferromagnetic critical behavior is controlled by the clean 2D Ising fixed point, consistent with the fact that random-mass disorder is (marginally) Harris irrelevant at that fixed point~\cite{dotsenko1983,zhu2015}. For sufficiently strong disorder, however, the clean critical behavior gives way to critical behavior controlled by a zero-temperature disordered fixed point (spin-glass critical point) via an intervening disordered multicritical point, the Nishimori point~\cite{nishimori1980,*nishimori1981,honecker2001}.

Coming back to the RG flow of the random-mass chiral XY GNY model~\cite{Yerzhakov2018}, depending on the number of fermion flavors (see Sec.~\ref{sec:model}) the disordered critical point (DFP$_2$) is found to be either a standard sink-type fixed point, as illustrated in Fig.~\ref{fig:RG}(d), or a fixed point of stable-focus type. In the latter case, RG trajectories asymptotically spiral towards the fixed point, implying oscillatory corrections to scaling. Stable-focus fixed points have been found before in replica RG studies of both classical~\cite{aharony1975} and quantum~\cite{Khmelnitskii1978,Dorogovtsev1980,Boyanovsky1982,boyanovsky1983,Lawrie1984,kirkpatrick1996} disordered systems, and are sometimes considered an artefact of perturbative replica-based RG. However, such flows cannot be ruled out as a matter of principle, since DFPs are in general non-unitary, and real, non-unitary, scale-invariant quantum field theories can have pairs of scaling fields with complex-conjugate dimensions~\cite{aharony2018,gorbenko2018}. Furthermore, oscillations in scaling laws, characteristic of spiraling or cyclic RG flows, have also been found in numerical studies of disordered holographic models~\cite{hartnoll2016} which rely neither on the replica trick nor on perturbation theory (in either the interaction or disorder strengths). A recent Monte Carlo study of classically frustrated 3D Heisenberg antiferromagnets also supports the existence of a stable-focus critical point (in this case, in a clean system)~\cite{nagano2019}.

In the present work, we extend the study of Ref.~\cite{Yerzhakov2018} in two directions. First, in Ref.~\cite{Yerzhakov2018}, only short-range correlated (or equivalently at long distances, uncorrelated) random-mass disorder was considered. Here we additionally consider random-mass disorder with correlations between two spatial points $\b{x},\b{x}'$ that decay asymptotically as a power law, $\sim|\b{x}-\b{x}'|^{-\alpha}$, with $\alpha<d$. (For $\alpha>d$, the correlations are short range, as the disorder correlation function in momentum space remains finite in the long-wavelength limit.) A clean critical point with correlation length exponent $\nu_\text{CFP}$ is perturbatively stable against such long-range correlated disorder if $\nu_\text{CFP} > 2/\min(d,\alpha)$~\cite{Weinrib1983}; this type of disorder thus generally has a stronger effect at phase transitions than uncorrelated disorder. Second, Ref.~\cite{Yerzhakov2018} only studied the chiral XY GNY model. Here, we perform a comprehensive study of the effect of random-mass disorder in the three standard families of critical GNY models: the chiral Ising, XY, and Heisenberg models~\cite{Rosenstein1993,Zerf2017}, fermionic analogs of the Ising, XY, and Heisenberg Wilson-Fisher universality classes, respectively. As we briefly review in Sec.~\ref{sec:model}, these chiral GNY models describe a variety of QCPs in condensed matter systems~\cite{boyack2020}.

Our main results are summarized as follows. For the chiral Ising GNY model, we find new disordered multicritical points, and for the chiral XY and Heisenberg GNY models, new disordered critical and multicritical points. As in Ref.~\cite{Yerzhakov2018}, some of the disordered QCPs found exhibit usual sink-type RG flows, while others are of stable-focus type. We also explore how the structure of the RG flow on the critical hypersurface evolves upon tuning RG-invariant system parameters, here the number $N$ of fermion flavors and the exponent $\alpha$ describing disorder correlations. We are particularly interested in bifurcations of these RG flows~\cite{gukov2017}, where the number or stability properties of fixed points suddenly change as a function of $N$ and $\alpha$, called control parameters in bifurcation theory. We find and analyze instances of the saddle-node bifurcation, also known as the fixed-point annihilation scenario~\cite{Kaplan2009}, at which a repulsive fixed point and an attractive fixed point coalesce and disappear into the complex plane. This type of bifurcation appears or has been argued to appear in RG flows in a variety of problems of current interest in both high-energy physics~\cite{kubota2001,kaveh2005,gies2006,Kaplan2009,braun2011,herbut2016,gracey2018,gorbenko2018} and condensed matter physics/statistical mechanics~\cite{herbut2014,janssen2015,*janssen2016,*janssen2017,nahum2015,wang2017,gorbenko2018b,serna2019,ihrig2019,ma2019,nahum2019}. The characteristic phenomenology associated with it includes Berezinskii-Kosterlitz-Thouless/Miransky scaling, walking/pseudo-critical behavior, and weakly first-order transitions. In our particular problem, it manifests itself in the existence of an anomalously (i.e., exponentially) large length scale $L_*$ that governs the crossover between two distinct universality classes of critical behavior. In much previous work, the saddle-node bifurcation is tuned by a parameter such as space(time) dimensionality $d$ or the integer number $N$ of components of a fermionic or bosonic field, and thus cannot be approached continuously in practice. Here, for fixed $d$ and $N$ the bifurcation can be approached by continuously tuning the exponent $\alpha$ for disorder correlations.

Besides the saddle-node bifurcation, we also discover instances of more exotic bifurcations~\cite{gukov2017}: the transcritical bifurcation, at which two fixed points exchange their stability properties without annihilating, and the supercritical Hopf (or Poincar\'e-Andronov-Hopf) bifurcation~\cite{Marsden1976}. The latter is a bifurcation at which a stable-focus QCP loses its stability by giving birth to a stable limit cycle, which then controls the asymptotic critical behavior. A possibility first considered by Wilson~\cite{wilson1971}, stable RG limit cycles lead to log-periodic scaling behavior~\cite{Veytsman1993}, i.e., discrete scale invariance (as opposed to log-periodic behavior of {\it corrections} to scaling at stable-focus points). Hopf bifurcations in RG flows were found in classical disordered $O(n)$ models~\cite{Weinrib1983,Athorne1985,*Athorne1986}, but only the subcritical Hopf bifurcation~\cite{Marsden1976} was found, where an unstable-focus fixed point becomes stable and gives birth to an {\it unstable} limit cycle. As a result, the models studied in Refs.~\cite{Weinrib1983,Athorne1985,*Athorne1986} did not exhibit log-periodic critical scaling behavior in the long-distance limit.

The rest of the paper is structured as follows. In Sec.~\ref{sec:model}, we briefly describe the chiral GNY models with long-range correlated random-mass quenched disorder. In Sec.~\ref{sec:RG}, we describe the perturbative RG scheme used to derive beta functions on the critical hypersurface. By contrast with Ref.~\cite{Yerzhakov2018}, where the double epsilon~\cite{Dorogovtsev1980,Boyanovsky1982,boyanovsky1983,Lawrie1984} expansion was sufficient to tame RG flows in the presence of uncorrelated disorder, here we use a controlled {\it triple} epsilon expansion~\cite{DeCesare1994} at one-loop order that allows us to tame the flow of both interaction and correlated disorder strengths. In Sec.~\ref{sec:FPs}, we investigate the fixed points of the RG beta functions derived in Sec.~\ref{sec:RG}, focusing on DFPs and analyzing their linear stability. We compute critical exponents and anomalous dimensions at all DFPs. In Sec.~\ref{sec:RGflows}, we discuss qualitative features of the RG flow, including various bifurcations that occur under changes of the control parameters $N$ and $\alpha$, and their consequences for critical properties. We conclude in Sec.~\ref{sec:conclusion} with a summary of our main results and a few directions for further research. Three appendices (App.~\ref{app:Z}-\ref{app:LimitCycleScaling}) contain the details of some calculations.

\section{The random-mass GNY models}
\label{sec:model}

Our starting point is the family of chiral $O(n)$ GNY models in 2+1 dimensions at zero temperature, described by the Euclidean action:
\begin{align}\label{S}
S=\int d^2\b{x}\,d\tau \left( \mathcal{L}_\phi+\mathcal{L}_\psi+\mathcal{L}_{\psi \phi} \right),
\end{align}
where $\b{x}$ denotes spatial coordinates, and $\tau$ is imaginary time. The model consists of a real $n$-component scalar field $\b{\phi}=(\phi^1,\ldots,\phi^n)$, the order parameter, governed by the Lagrangian:
\begin{align}\label{Lphi}
\c{L}_\phi=(\partial_\tau\b{\phi})^2+c_b^2(\nabla\b{\phi})^2+r\b{\phi}^2+\lambda^2(\b{\phi}^2)^2,
\end{align}
where $\b{\phi}^2=\b{\phi}\cdot\b{\phi}=\sum_{i=1}^n(\phi^i)^2$. It is coupled to a Dirac fermion field $\psi$, described by the Lagrangian:
\begin{align}\label{Lpsi}
\mathcal{L}_\psi =  i \overline{\psi} (\gamma_0 \partial_\tau + c_f \b{\gamma} \cdot \nabla) \psi.
\end{align}
The scalar mass squared $r$ in Eq.~(\ref{Lphi}) tunes the model through criticality: $r<0$ gives a phase with spontaneously broken $O(n)$ symmetry, $r>0$ is the symmetric phase, and $r=0$ is the critical point. The parameter $\lambda^2$ describes self-interactions of the order parameter. We define the Dirac adjoint in Eq.~(\ref{Lpsi}) as $\overline{\psi} = -i \psi^\dagger \gamma_0$. We denote $\b{\gamma}=(\gamma_1,\gamma_2)$, and $\gamma_\mu$, $\mu=0,1,2$ are Hermitian Dirac matrices obeying the $SO(3)$ Clifford algebra $\{\gamma_\mu,\gamma_\nu\}=2\delta_{\mu\nu}$. In the ordinary GNY model, Lorentz invariance (exact or emergent at criticality~\cite{roy2016}) demands that the fermion $c_f$ and boson $c_b$ velocities be equal, but in the presence of quenched disorder, to be introduced below, the ratio $c=c_f/c_b$ will flow under RG transformations.

We perform perturbative calculations near four dimensions at one-loop order in the context of a particular epsilon-expansion scheme to be explained below, but we are ultimately interested in (2+1)D physics. As is customary for these types of problems (see, e.g., Ref.~\cite{Zerf2017}), we adopt a naive dimensional-regularization prescription according to which all Dirac matrices anticommute~\cite{chanowitz1979} and spinor traces over products of an odd number of Dirac matrices vanish~\cite{DREG3}. In addition to a spinor index, the field $\psi$ carries a flavor index. With the dimensional-regularization prescription just mentioned, perturbative results only depend on the total number of (complex) fermionic degrees of freedom, i.e., the dimension of the chosen representation of the Dirac algebra, times the number of flavors. We will present our results in terms of the number $N$ of flavors of two-component Dirac fermions (i.e., the number of linear band crossing points at the Fermi level in a condensed matter system), but they can alternatively be interpreted as pertaining to $N_f=N/2$ flavors of four-component Dirac fermions when $N$ is even. 

We consider the cases $n=1,2,3$, corresponding to the chiral Ising, XY, and Heisenberg GNY models, respectively~\cite{Rosenstein1993,Zerf2017}. The form of the Yukawa coupling $\c{L}_{\psi\phi}$ in Eq.~(\ref{S}) differs in each case. In the chiral Ising GNY model~\cite{zinn-justin1991}, a single real scalar $\phi$ couples to the fermion mass $i\overline{\psi}\psi$,
\begin{align}\label{YukIsing}
\mathcal{L}_{\psi \phi}^\text{Ising}= i h  \phi \overline{\psi}  \psi,
\end{align}
with coupling strength $h$. The Yukawa coupling in the chiral XY GNY model can be formulated in different but equivalent ways, depending on the choice of spinor representation. In the four-component representation, the Yukawa coupling can be written as a coupling to both the ordinary mass $i\overline{\psi}\psi$ and an axial mass $\overline{\psi}\gamma_5\psi$,
\begin{align}
\c{L}_{\psi\phi}^\text{XY}=ih\overline{\psi}(\phi^1+i\gamma_5\phi^2)\psi,
\end{align}
and is equivalent to the Nambu--Jona-Lasinio model~\cite{nambu1961}. Here, one utilizes a four-dimensional representation $\gamma_\mu$, $\mu=0,1,2,3$ of the $SO(4)$ Clifford algebra, and $\gamma_5=\gamma_0\gamma_1\gamma_2\gamma_3$. In a different spinor representation~\cite{SpinorRep}, the model can be written as a coupling to a Majorana mass,
\begin{align}\label{YukXY}
\c{L}_{\psi\phi}^\text{XY}=\frac{h}{2}(\phi^*\psi^Ti\gamma_2\psi+\mathrm{H.c.}),
\end{align}
where the $O(2)$ order parameter $\b{\phi}=(\phi^1,\phi^2)$ is expressed as a complex scalar field $\phi=\phi^1+i\phi^2$. Finally, the Yukawa coupling in the chiral Heisenberg GNY model is:
\begin{align}\label{YukHeis}
\c{L}_{\psi\phi}^\text{Heis}=ih\b{\phi}\cdot\overline{\psi}\bsigma\psi,
\end{align}
where $\bsigma=(\sigma_1,\sigma_2,\sigma_3)$ forms a spin-1/2 representation of the $SU(2)$ algebra.

For different values of $N$, the $O(n)$ GNY models introduced above describe a variety of quantum phase transitions in (2+1)D condensed matter systems~\cite{boyack2020}. For $N=4$ (spinful fermions) and $N=2$ (spinless fermions), the chiral Ising GNY model ($n=1$) describes a transition from a Dirac semimetal to an insulator with charge-density-wave order on the honeycomb lattice~\cite{herbut2006}. For $N=1$, the model describes a ferromagnetic transition on the surface of a 3D topological insulator~\cite{Xu2010}. For $N=1/2$, which can be interpreted as a model containing a single flavor of two-component Majorana fermions, the model describes the time-reversal symmetry-breaking transition on the surface of a 3D topological superconductor~\cite{Grover2014}, which exhibits an emergent $\c{N}=1$ supersymmetry~\cite{sonoda2011,Grover2014,fei2016}. Turning to the chiral XY GNY model ($n=2$), the cases $N=4$ and $N=2$ describe a quantum phase transition from a Dirac semimetal (spinful or spinless, respectively) to an insulator with Kekul\'e valence-bond-solid (VBS) order on the honeycomb lattice~\cite{hou2007,Roy2013}, or to an insulator with columnar VBS order on the $\pi$-flux square lattice~\cite{zhou2018}. The spontaneously broken symmetries in those examples are discrete $\mathbb{Z}_3$ and $\mathbb{Z}_4$ point group symmetries, respectively, but those anisotropies are irrelevant perturbations at the $O(2)$-symmetric GNY fixed point, at least in the large-$N$ limit~\cite{li2017,zerf2020}. However, in those VBS realizations of chiral XY GNY criticality, spatial randomness necessarily couples linearly to the VBS order parameter: it thus acts as random-field disorder, which destroys the $d=2$ critical point~\cite{RandomField}. Alternatively, the chiral XY GNY model also describes a semimetal-superconductor transition in a system with $N$ two-component Dirac fermions ($N=4$ for spinful fermions on the honeycomb lattice~\cite{Roy2013}), in which case the $U(1)\cong SO(2)$ symmetry is exact and random-field disorder is forbidden by conservation of particle number. For $N=1$, the model describes a superconducting transition on the surface of a 3D topological insulator, and exhibits an emergent $\mathcal{N}=2$ supersymmetry~\cite{Grover2014,ponte2014,Roy2013,Zerf2016,fei2016,witczak-krempa2016}. Finally, for $N=4$ the chiral Heisenberg GNY model ($n=3$) describes the transition from a Dirac semimetal to an insulator with antiferromagnetic spin-density-wave order on the honeycomb lattice~\cite{herbut2006}.

We model quenched random-mass disorder by randomness in the scalar mass squared, $r(\b{x})=r_0+\delta r(\b{x})$, where $\delta r(\b{x})$ is a Gaussian random variable of zero mean and correlation function~\cite{Weinrib1983}:
\begin{align}\label{DisCorr}
\overline{\delta r(\b{x})\delta r(\b{x}')}\propto\Delta\delta(\b{x}-\b{x}')+\frac{v}{|\b{x}-\b{x}'|^\alpha},
\end{align}
where $\overline{\cdots}$ denotes disorder averaging. (Random-mass disorder that couples directly to fermions is perturbatively irrelevant in the epsilon-expansion scheme we utilize, as we explain in more detail in Sec.~\ref{sec:RG}.) The uniform part $r_0$ is the tuning parameter for the transition, and $\Delta$ and $v$ are the short-range and long-range correlated disorder strengths, respectively. Even when considering initial conditions for the RG with only long-range correlated disorder, $\Delta=0$, short-range correlated disorder is generated perturbatively already at one-loop order, see Eq.~(\ref{bD}), and should be kept in the space of couplings. By contrast, long-range correlated disorder cannot be generated perturbatively from short-range correlated disorder, see Eq.~(\ref{bv}). We use the replica trick to average over disorder~\cite{QPT}, which induces an effective two-body interaction,
\begin{align}\label{Sdis}
S_\text{dis}=-\frac{\Delta}{2} \sum_{ab}\int d^2\b{x}\,d\tau\,d\tau'\,
\b{\phi}_a^2(\b{x},\tau)\b{\phi}_b^2(\b{x},\tau') 
 -\frac{v}{2} \sum_{ab}\int d^2\b{x}\,d^2\b{x}'\,d\tau\,d\tau'\, \frac{
\b{\phi}_a^2(\b{x},\tau)\b{\phi}_b^2(\b{x},\tau')}{|\b{x}-\b{x}'|^\alpha},
\end{align}
where $a,b=1,\ldots,m$ are replica indices, and the replica limit $m\rightarrow 0$ is to be taken at the end of the calculation. As for the superfluid-Mott glass transition~\cite{weichman2008}, randomness in the scalar mass squared preserves the exact particle-hole symmetry of the clean GNY action (\ref{S}).

\section{RG in the triple epsilon expansion}
\label{sec:RG}

We first briefly recapitulate the idea of the double epsilon expansion for QCPs perturbed by quenched short-range correlated disorder, first focusing on the purely bosonic random-mass $O(n)$ vector model~\cite{Dorogovtsev1980,Boyanovsky1982,boyanovsky1983,Lawrie1984}. In $d=4-\epsilon$ spatial and $\epsilon_\tau$ imaginary time dimensions, the order parameter field $\b{\phi}$ has engineering dimension $\Delta_\phi=(2-\epsilon+\epsilon_\tau)/2$. The couplings $\lambda^2$ and $\Delta$ thus have mass dimension $\epsilon-\epsilon_\tau$ and $\epsilon$, respectively, and a controlled perturbative RG analysis can be performed by treating $\epsilon$ and $\epsilon_\tau$ as small parameters. For $n>1$, a stable DFP with $\lambda^2_*\sim\mathcal{O}(\epsilon,\epsilon_\tau)$, $\Delta_*\sim\mathcal{O}(\epsilon,\epsilon_\tau)$ on the critical hypersurface $r=0$ is found at one-loop order, with critical exponents~\cite{NoteBCFP}:
\begin{align}
\nu&=\frac{1}{2}+\frac{3n\epsilon+(2n+4)\epsilon_\tau}{32(n-1)},\label{BCnu}\\
z&=1+\frac{(4-n)\epsilon+(2n+4)\epsilon_\tau}{16(n-1)}.\label{BCz}
\end{align}
For $n=2$, and extrapolating $\epsilon_\tau$ to 1 and $\epsilon$ to 2 or 1, relevant to the boson superfluid-Mott glass transition in (2+1)D and (3+1)D, respectively, one obtains exponents in reasonable agreement with those found in numerical Monte Carlo (MC) simulations (Table~\ref{tab:SFMG}). Ref.~\cite{Yerzhakov2018} observed that the double epsilon expansion can also be applied to short-range correlated random-mass GNY models: the fermion field $\psi$ has engineering dimension $\Delta_\psi=(3-\epsilon+\epsilon_\tau)/2$, thus the Yukawa coupling $h$ has mass dimension $(\epsilon-\epsilon_\tau)/2$ and can also be treated perturbatively. To the difference of the bosonic model, however, one must enlarge the space of running couplings to include the relative velocity $c=c_f/c_b$, and ensure that the beta function for this parameter also vanishes at the DFP. (For a disordered system with a single field, the flow of the velocity, e.g., $c_b$ for the bosonic $O(n)$ model, can be absorbed in the definition of $z$, provided the disorder strength flows to a fixed-point value $\Delta_*$~\cite{narovlansky2018,aharony2018}.) Random-mass disorder will also generally couple to the fermionic sector of the GNY models, the most relevant coupling being a random coupling to fermion bilinears. For Gaussian disorder, the resulting disorder-induced two-body coupling has mass dimension $-2+\epsilon$, and is thus strongly irrelevant in the epsilon expansion.

\begin{table}[t]
\begin{tabular}{|l||c|c|c|}
\hline 
 & MC~\cite{vojta2016,crewse2018} & $\c{O}(\epsilon,\epsilon_\tau)$ & $\c{O}(1/n)$~\cite{goldman2020}
 \tabularnewline
\hline 
\hline 
$\nu$, (2+1)D & 1.16(5) & 1.125 & 1 \\
\hline 
$z$, (2+1)D & 1.52(3) & 1.75 & 1.54 \\
\hline
\hline
$\nu$, (3+1)D & 0.90(5) & 0.9375 & --- \\
\hline
$z$, (3+1)D & 1.67(6) & 1.625 & --- \\
\hline
\end{tabular}
\caption{Critical exponents for the boson superfluid-Mott glass transition.}\label{tab:SFMG}
\end{table}

In the presence of long-range correlated disorder, we see from Eq.~(\ref{Sdis}) that the coupling constant $v$ has mass dimension $4-\alpha$ at the Gaussian fixed point. While for generic $\alpha<d<4$ this coupling is strongly relevant, if we set $\alpha=4-\delta$ and treat $\delta$ as a small parameter long-range correlated disorder is only slightly relevant and can be treated perturbatively~\cite{Weinrib1983}. (Fermionic disorder of the type discussed above but with long-range correlations~\cite{Fedorenko2012,Dudka2016} would have mass dimension $-2+\delta$ and is still irrelevant.) This forms the basis of a triple expansion in $\epsilon,\epsilon_\tau,\delta$~\cite{DeCesare1994}, which thus far has only been applied to bosonic systems. Below we employ this triple epsilon expansion to study the GNY models with both short-range and long-range correlated random-mass disorder.

In the presence of three epsilon-like parameters, the nature of the RG fixed points and their stability depend on two ratios, e.g., $\epsilon/\epsilon_\tau$ and $\delta/\epsilon_\tau$. We restrict our consideration to $\epsilon/\epsilon_\tau=2$, which in the limit $\epsilon_\tau\rightarrow 1$ corresponds to (2+1)D systems. Regarding the $\delta/\epsilon_\tau$ ratio, we consider the range $0<\delta/\epsilon_\tau<4$. For $\delta<0$, long-range correlated disorder is irrelevant at the Gaussian fixed point, and for $\delta/\epsilon_\tau>4$, the long-range disorder correlations (\ref{DisCorr}) with $\alpha=4-\delta$ would have the unphysical feature of increasing rather than decaying with distance in the limit $\epsilon_\tau\rightarrow 1$.

\subsection{Bare vs renormalized actions}

We now outline the basic steps of the RG procedure using as example the chiral XY GNY model studied in Ref.~\cite{Yerzhakov2018}, but with long-range correlated disorder (\ref{Sdis}). For the chiral Ising and Heisenberg GNY models, the number of components of the order parameter and the form of the Yukawa coupling change [see Eqs.~(\ref{YukIsing}-\ref{YukHeis})], but the relations (\ref{rel}) between bare and renormalized couplings, and the formal expressions (\ref{betac2}-\ref{betar}) for the beta functions in terms of the anomalous dimensions (\ref{Zi}), remain the same.

As in Refs.~\cite{boettcher2016,mandal2018}, we rescale the time coordinate as well as the fermion and boson fields, and redefine the couplings in the action (\ref{S}-\ref{Sdis}), to eliminate the velocities $c_f$ and $c_b$ in favor of the dimensionless ratio $c^2=(c_f/c_b)^2$, which then appears in front of the time derivative term for the boson field~\footnote{Alternatively, one can define fermionic $z_\psi$ and bosonic $z_\phi$ dynamic critical exponents from the flow of the velocities $c_f$ and $c_b$, respectively, which leads to renormalized dispersions $\omega_\psi(p)\sim p^{z_\psi}$, $\omega_\phi(p)\sim p^{z_\phi}$. The existence of a fixed point for $\beta_{c^2}$ then signifies that those exponents are in fact the same at criticality, $z_\psi=z_\phi=z$.}. The replicated bare action for the random-mass chiral XY GNY model is then:
\begin{align}\label{Sbare}
S_B=&\sum_a\int d^d\b{x}_B\,d^{\epsilon_\tau}\tau_B\biggl(i\overline{\psi}_{a,B}(\gamma_0\partial_{\tau_B}+\b{\gamma}\cdot\nabla_B)\psi_{a,B}
+\phi_{a,B}^*(-c^2_B\partial_{\tau_B}^2-\nabla_B^2+r)\phi_{a,B} \nn\\
&\hspace{35mm}+  \lambda_B^2|\phi_{a,B}|^4+\frac{h_B}{2}(\phi_{a,B}^*\psi_{a,B}^Ti\gamma_2\psi_{a,B}+\text{H.c.})\biggr)\nn\\
&-\frac{\Delta_B}{2}\sum_{ab}\int d^d\b{x}_B\,d^{\epsilon_\tau}\tau_B\,d^{\epsilon_\tau}\tau_B'
|\phi_{a,B}|^2(\b{x}_B,\tau_B)|\phi_{b,B}|^2(\b{x}_B,\tau_B')\nn \\
&-\frac{v_B}{2} \sum_{ab}\int d^d\b{x}_B\,d^d\b{x}'_B\,d^{\epsilon_\tau}\tau_B\,d^{\epsilon_\tau}\tau'_B\, \frac{
|\phi_{a,B}|^2(\b{x}_B,\tau_B)|\phi_{b,B}|^2(\b{x}'_B,\tau'_B)}{ |\b{x}_B-\b{x}'_B|^\alpha},
\end{align}
where $a,b=1,\ldots,m$ are replica indices, and the corresponding renormalized action is:
\begin{align}\label{Sren}
S=&\sum_a\int d^d\b{x}\,d^{\epsilon_\tau}\tau\biggl(i\overline{\psi}_a(Z_1\gamma_0\partial_\tau+Z_2\b{\gamma}\cdot\nabla)\psi_a
+\phi_a^*(-Z_3c^2\partial_\tau^2-Z_4\nabla^2+Z_rr\mu^2)\phi_a \nn\\
&\hspace{30mm}+Z_5\lambda^2\mu^{\epsilon-\epsilon_\tau}|\phi_a|^4
+Z_6\frac{h}{2}\mu^{(\epsilon-\epsilon_\tau)/2}(\phi_a^*\psi_a^Ti\gamma_2\psi_a+\text{H.c.})\biggr)\nn\\
&-Z_7\frac{\Delta}{2}\mu^\epsilon\sum_{ab}\int d^d\b{x}\,d^{\epsilon_\tau}\tau\,d^{\epsilon_\tau}\tau'\,
|\phi_a|^2(\b{x},\tau)|\phi_b|^2(\b{x},\tau') \nn\\
&-Z_8\frac{v}{2} \mu^\delta \sum_{ab}\int d^d\b{x}\,d^d\b{x}'\,d^{\epsilon_\tau}\tau\,d^{\epsilon_\tau}\tau'\, \frac{
|\phi_a|^2(\b{x},\tau)|\phi_b|^2(\b{x}',\tau')}{ |\b{x}-\b{x}'|^\alpha},
\end{align}
where $\mu$ is a renormalization scale. Due to the anisotropy between space and time, we set $\b{x}_B=\b{x}$ and $\tau_B=\eta\tau$, and matching the bare and renormalized kinetic terms for the fermion we find that $\eta=Z_{2}/Z_{1}$. Defining the anomalous dimensions:
\begin{align}\label{Zi}
\gamma_{i}=\mu\frac{d\ln Z_{i}}{d\mu},\,i=1,\ldots,8, r,
\end{align}
we find that the dynamic critical exponent $z=\mu(d\ln\tau/d\mu)$~\cite{Thomson2017} is given by:
\begin{align}
z=1+\gamma_{1}-\gamma_{2}.
\end{align}
The fermion and boson fields are multiplicatively renormalized,
\begin{align}
\psi_{a,B}(\b{x}_B,\tau_B)=\sqrt{Z_\psi}\psi_a(\b{x},\tau),\hspace{5mm}
\phi_{a,B}(\b{x}_B,\tau_B)=\sqrt{Z_\phi}\phi_a(\b{x},\tau),
\end{align}
and the fermion and boson anomalous dimensions, $\eta_\psi=\mu(d\ln Z_{\psi}/d\mu)$ and $\eta_\phi=\mu(d\ln Z_{\phi}/d\mu)$, are given by:
\begin{align}
\eta_\psi=\gamma_2+\epsilon_\tau(z-1),\hspace{5mm}
\eta_\phi=\gamma_4+\epsilon_\tau(z-1).
\end{align}
Comparing Eqs.~(\ref{Sbare}) and (\ref{Sren}), we obtain relations between the bare and (dimensionless) renormalized couplings,
\begin{align}\label{rel}
c^2&=Z_{3}^{-1}Z_{4}\left(\frac{Z_{1}}{Z_{2}}\right)^2 c_{B}^2,\hspace{5mm}
\lambda^2=\mu^{-(\epsilon-\epsilon_\tau)}\left(\frac{Z_{1}}{Z_{2}}\right)^{\epsilon_\tau}Z_{4}^2Z_{5}^{-1}\lambda_{B}^2,\hspace{5mm}
h^2=\mu^{-(\epsilon-\epsilon_\tau)}\left(\frac{Z_{1}}{Z_{2}}\right)^{\epsilon_\tau}Z_{2}^2Z_{4}Z_{6}^{-2}h_{B}^2, 
\nn\\
\Delta&=\mu^{-\epsilon}Z_{4}^2Z_{7}^{-1}\Delta_{B}, \hspace{5mm}
v=\mu^{-\delta}Z_{4}^2Z_{8}^{-1}v_{B}, \hspace{5mm}
r=\mu^{-2}Z_4Z_r^{-1}r_B.
\end{align}
Using the fact that the bare couplings do not depend on the renormalization scale $\mu$, we find the RG beta functions $\beta_g\equiv\mu(dg/d\mu)$, $g\in\{c^2,\lambda^2,h^2,\Delta,v\}$, to be:
\begin{align}
\beta_{c^2}&=(2\gamma_{1}-2\gamma_{2}-\gamma_{3}+\gamma_{4})c^2,\label{betac2}\\
\beta_{\lambda^2}&=\bigl(-(\epsilon-\epsilon_\tau)+2\gamma_{4}-\gamma_{5}+\epsilon_\tau(\gamma_{1}-\gamma_{2})\bigr)
\lambda^2,\label{betal2formal}\\
\beta_{h^2}&=\bigl(-(\epsilon-\epsilon_\tau)+2(\gamma_{2}-\gamma_{6})+\gamma_{4}+\epsilon_\tau(\gamma_{1}-\gamma_{2}) \bigr)h^2,\label{betah2formal}\\
\beta_{\Delta}&=(-\epsilon+2\gamma_{4}-\gamma_{7})\Delta,\label{betaDelta}\\
\beta_{v}&=(-\delta+2\gamma_{4}-\gamma_{8})v,\label{betav} \\
\beta_r&=(-2+\gamma_4-\gamma_r)r. \label{betar}
\end{align}
From Eq.~(\ref{betar}), we find the inverse correlation length exponent~\cite{Zinn-JustinBook},
\begin{align}
\nu^{-1}=2-\gamma_4+\gamma_r.
\end{align}

\subsection{Renormalization constants}

\begin{figure}[t]
\includegraphics[width=0.65\columnwidth]{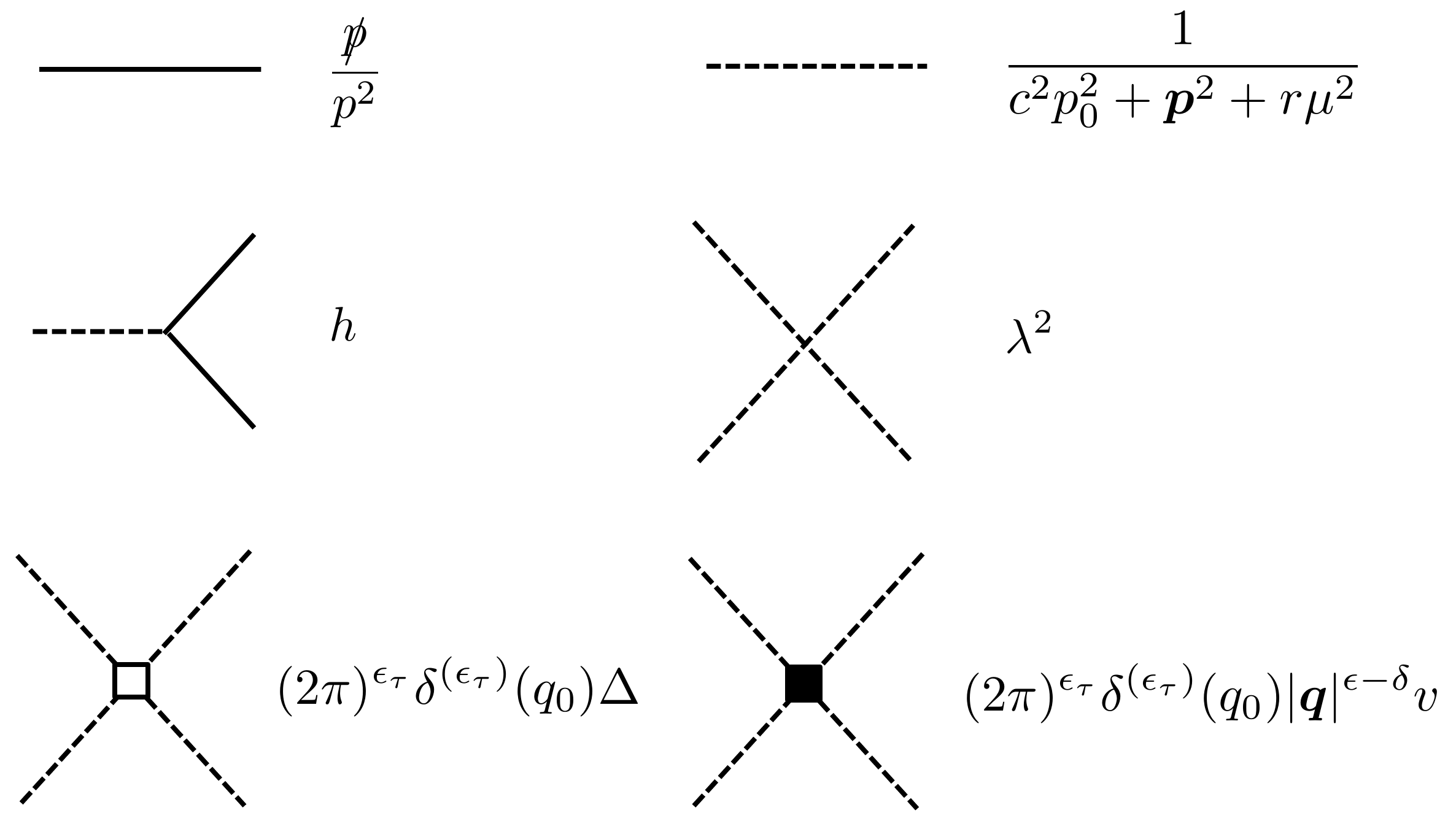}
\caption{Schematic momentum-space Feynman rules for the random-mass GNY models, omitting fermion flavor, $O(n)$, and replica indices. Solid line: fermion propagator, dashed line: boson propagator. Here $p=(p_0,\b{p})$ is the momentum of a propagator line, with $\slashed{p}=\gamma_\mu p_\mu$, and $q=(q_0,\b{q})$ is the momentum transfer in a boson four-point vertex.}
\label{fig:feynrules}
\end{figure}

\begin{figure}[t]
\includegraphics[width=0.8\columnwidth]{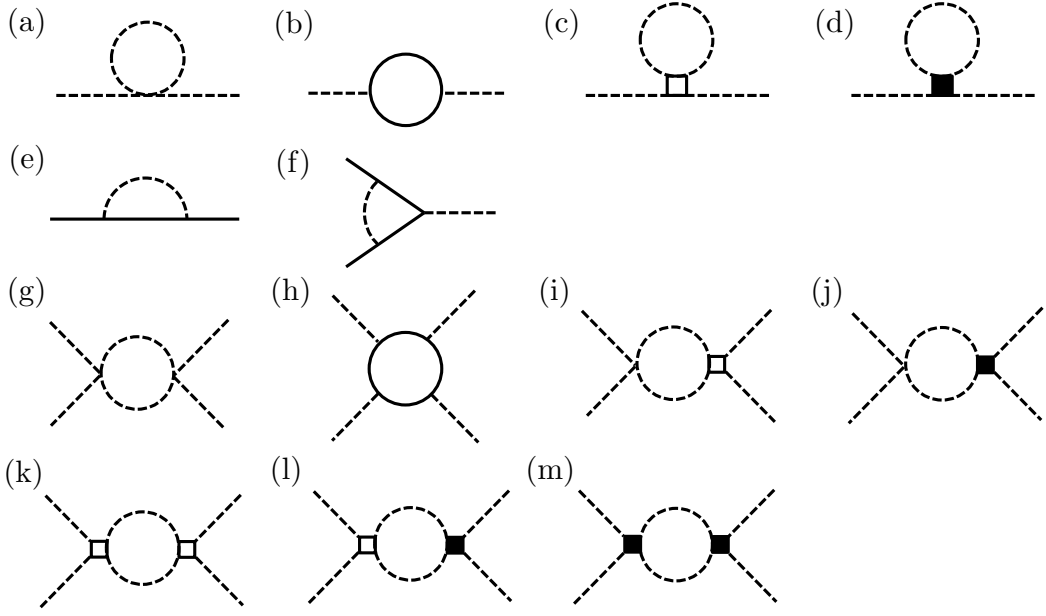}
\caption{Schematic one-loop Feynman diagrams for the random-mass GNY models. Renormalization of (a,b,c,d) the boson two-point function; (e) the fermion two-point function; (f) the Yukawa vertex $h$; (g,h,i,j) the boson self-interaction vertex $\lambda^2$; (i,k,l,m) the short-range correlated disorder vertex $\Delta$; and (j,l,m) the long-range correlated disorder vertex $v$.}
\label{fig:diagrams}
\end{figure}

We calculate the renormalization constants $Z_i$, $i=1,\dots,8,r$ at one-loop order in the modified minimal subtraction ($\overline{\text{MS}}$) scheme with dimensional regularization in $4-\epsilon$ space and $\epsilon_\tau$ time dimensions. The relevant Feynman rules and diagrams are shown schematically in Figs.~\ref{fig:feynrules} and \ref{fig:diagrams}, respectively. The fermion and boson propagators are given by:
\begin{align}
G_{ab}^{IJ}(p)&=\langle\psi_a^I(p)\overline{\psi}_b^J(p)\rangle=\delta_{ab}\delta^{IJ}\frac{\slashed{p}}{p^2},\\ 
D_{ab}^{ij}(p)&=\langle\phi^{i}_a(p)\phi^{j}_b(-p)\rangle=\delta_{ab} \delta^{ij}\frac{1}{c^2p_0^2+\b{p}^2+r\mu^2},\label{BosPropag}
\end{align}
where $I,J=1,\ldots,N$ and $i,j=1,\ldots,n$ are fermion flavor and $O(n)$ indices, respectively, and $\slashed{p}=\gamma_\mu p_\mu$.

\begin{figure}[t]
\includegraphics[width=0.5\textwidth]{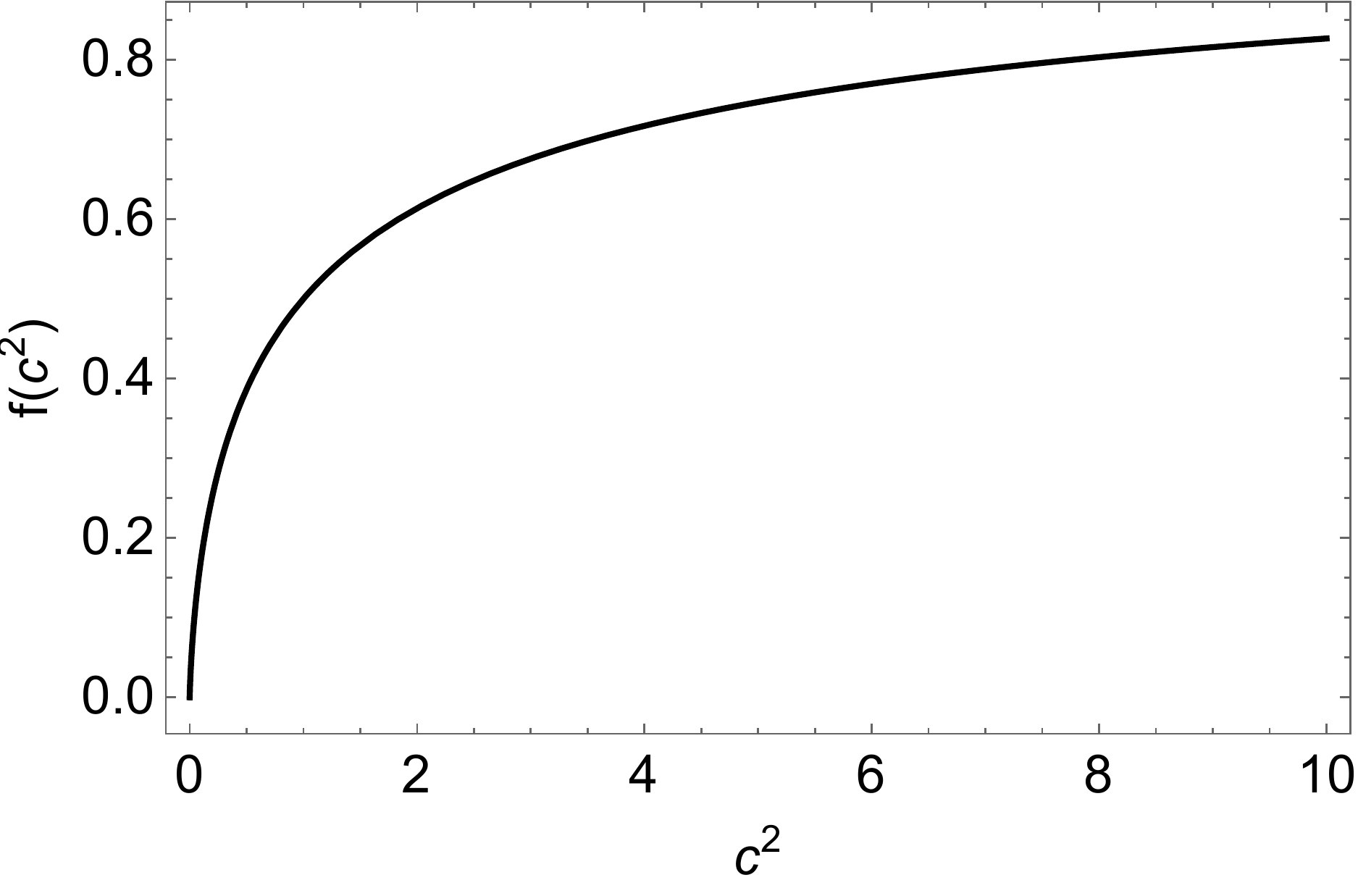}
\caption{Plot of $f(c^2)$ in Eq.~(\ref{f}), with $c^2=(c_f/c_b)^2$ the velocity ratio squared; $f(0)=0$, $f(1)=\frac{1}{2}$, and $f(\infty)=1$.}
\label{fig:f}
\end{figure}

For the chiral XY GNY model ($n=2$), the diagrams in the clean limit or containing only short-range correlated disorder vertices were already computed in Ref.~\cite{Yerzhakov2018}; these results are also easily adapted to $n=1$ and $n=3$. The new diagrams containing long-range correlated disorder vertices are evaluated explicitly in Appendix~\ref{app:Z} for $n=1,2,3$. Unlike the standard epsilon expansion in $4-\epsilon$ dimensions, in the triple epsilon expansion one-loop diagrams contain simple poles not only in $\epsilon$, but also in $\epsilon-\epsilon_\tau$, $\delta$, and $2\delta - \epsilon$. We obtain the following renormalization constants:
\begin{align}
Z_{1}&=1- \frac{n h^2}{\epsilon-\epsilon_\tau}f(c^2),\label{Z1}\\
Z_{2}&=1- \frac{n h^2}{2(\epsilon-\epsilon_\tau)},\\
Z_{3}&=1-\frac{2\Delta}{\epsilon}-\frac{2v}{\delta} -\frac{Nh^2c^{-2}}{\epsilon-\epsilon_\tau},\\
Z_{4}&=1-\frac{Nh^2}{\epsilon-\epsilon_\tau},\\
Z_{5}&=1+\frac{2(n+8)\lambda^2}{\epsilon-\epsilon_\tau}-\frac{Nh^4\lambda^{-2}}{\epsilon-\epsilon_\tau}-\frac{12\Delta}{\epsilon}-\frac{12v}{\delta},\\
Z_{6}&=1+(2-n) \frac{h^2}{\epsilon-\epsilon_\tau},\label{Z6}\\
Z_{7}&=1+\frac{4(n+2)\lambda^2}{\epsilon-\epsilon_\tau}-\frac{8\Delta}{\epsilon}-\frac{12v}{\delta}-\frac{4v^2 \Delta^{-1}}{2\delta-\epsilon},\label{Z7}\\
Z_{8}&=1+\frac{4(n+2)\lambda^2}{\epsilon-\epsilon_\tau}-\frac{4\Delta}{\epsilon}-\frac{4v}{\delta},\label{Z8}\\
Z_{r}&=1+\frac{2(n+2)\lambda^2}{\epsilon-\epsilon_\tau}-\frac{2\Delta}{\epsilon}-\frac{2v}{\delta}.\label{Zr}
\end{align}
We have rescaled the couplings according to $g/(4\pi)^2\rightarrow g$, $g\in\{\lambda^2,h^2,\Delta,v, r\}$, and, as in Ref.~\cite{Yerzhakov2018}, we define the dimensionless function,
\begin{align}\label{f}
f(c^2)=\frac{c^2(c^2-1-\ln c^2)}{(c^2-1)^2},
\end{align}
plotted in Fig.~\ref{fig:f}. At one-loop order there is no renormalization of the Yukawa vertex for the chiral XY GNY model, i.e., the diagram in Fig.~\ref{fig:diagrams}(f) vanishes for $n=2$ [see Eq.~(\ref{Z6})], which is easily seen from the form (\ref{YukXY}) of the Yukawa coupling. We also see from the last term in Eq.~(\ref{Z7}) that short-range correlated disorder is generated at one-loop order from long-range correlated disorder, via the diagram in Fig.~\ref{fig:diagrams}(m). By contrast, long-range correlated disorder cannot be generated perturbatively from short-range correlated disorder.

\subsection{Beta functions and anomalous dimensions}

Using the chain rule,
\begin{align}\label{gammai}
\gamma_{i}=\frac{\mu}{Z_i}\frac{dZ_{i}}{d\mu}=\frac{1}{Z_{i}}\sum_g\frac{\partial Z_{i}}{\partial g}\beta_g,
\end{align}
for $i=1,\ldots,8,r$ and $g\in\{c^2,\lambda^2,h^2,\Delta,v,r\}$ in Eqs.~(\ref{betac2}-\ref{betar}), and expanding the beta functions to quadratic order in all couplings except $c^2$, we obtain:
\begin{align}
\beta_{c^2}&=-2(\Delta+v)c^2+h^2\big[N(c^2-1)+n c^2\left(2f(c^2)-1\right) \big],\label{bc2}\\
\beta_{\lambda^2}&=-(\epsilon-\epsilon_\tau)\lambda^2+2(n+8)\lambda^4+2Nh^2\lambda^2 
-Nh^4-12(\Delta+v)\lambda^2,\label{bl2}\\
\beta_{h^2}&=-(\epsilon-\epsilon_\tau)h^2+(N+4-n)h^4,\label{bh2}\\
\beta_\Delta&=-\epsilon\Delta+4(n+2)\lambda^2\Delta+2N h^2\Delta-8\Delta^2-12\Delta v-4v^2,\label{bD}\\
\beta_v&=-\delta v+4(n+2)\lambda^2v+2N h^2v-4\Delta v-4v^2.\label{bv}
\end{align}
We note that all poles in linear combinations of the small parameters $\epsilon,\epsilon_\tau,\delta$ properly cancel in the beta functions. Setting $\epsilon_\tau$ and the disorder couplings to zero, we find that Eqs.~(\ref{bl2}-\ref{bh2}) agree with the beta functions for the chiral $O(n)$ GNY models in the clean limit~\cite{Zerf2017}. When setting $n=2$ and $v=0$, Eqs.~(\ref{bc2}-\ref{bD}) reproduce our previous results for the chiral XY GNY model with short-range correlated disorder~\cite{Yerzhakov2018}. Finally, when turning off the Yukawa coupling, $h^2=0$, the beta functions (\ref{bl2},\ref{bD},\ref{bv}) with both short-range and long-range correlated disorder agree with those given in Refs.~\cite{Boyanovsky1982,boyanovsky1983,Lawrie1984,Weinrib1983,DeCesare1994}. We also note that the above beta functions are perturbative in the couplings $\lambda^2$, $h^2$, $\Delta$, and $v$, but exact in the dimensionless velocity ratio $c^2$.

The critical exponents $\nu^{-1}$, $z$, $\eta_\psi$, and $\eta_\phi$ are obtained by evaluating:
\begin{align}
\nu^{-1}&=2-Nh^2-2(n+2)\lambda^2+2(\Delta+v),\label{nuinv}\\
z&=1+\left(f(c^2)-\half\right)nh^2,\label{z}\\
\eta_{\psi}&=\frac{n}{2} h^2+\epsilon_\tau (z-1),\label{gammapsi}\\
\eta_{\phi}&=Nh^2+\epsilon_\tau (z-1),\label{gammaphi}
\end{align}
at RG fixed points $(c^2_*,\lambda^2_*,h^2_*,\Delta_*,v_*)$, i.e., common zeros of the set (\ref{bc2}-\ref{bv}) of beta functions. Since $h_*^2$ will be $\mathcal{O}(\epsilon,\epsilon_\tau)$ at one-loop order, as can already be seen from Eq.~(\ref{bh2}), for a consistent treatment we have to discard the $\epsilon_\tau (z-1)$ terms in the fermion and boson anomalous dimensions.

\section{Fixed points and critical exponents}
\label{sec:FPs}

In Sec.~\ref{sec:FixedPoints}, we discuss the fixed points of the flow equations (\ref{bc2}-\ref{bv}). Depending on their stability, which is analyzed in Sec.~\ref{sec:stability}, these are {\it bona fide} critical points (no relevant direction) or multicritical points (one or more relevant directions). Here, the number of relevant directions refers to the number of such directions on the critical hypersurface, since the tuning parameter $r$ for the transition (see Sec.~\ref{sec:model}) is a relevant direction at all fixed points. As mentioned in Sec.~\ref{sec:RG}, we fix $\epsilon=2\epsilon_\tau$, with the extrapolation $\epsilon_\tau\rightarrow 1$ corresponding to 2+1 dimensions. Throughout the paper, we evaluate quantities such as fixed-point couplings, RG eigenvalues, and critical exponents as a function of the control parameters $N\geqslant 1$ and $\delta=4-\alpha\in[0,4]$, where the latter parameter is to be understood as the ratio $\delta/\epsilon_\tau$ evaluated at $\epsilon_\tau=1$.

\subsection{Fixed points}
\label{sec:FixedPoints}

We denote the RG fixed points as five-component vectors $(c^2_*,\lambda^2_*,h^2_*,\Delta_*,v_*)$ in the space of running couplings. Starting with the CFPs ($\Delta_*=v_*=0$), these include Gaussian fixed points $(c_*^2,0,0,0,0)$ and the $O(n)$ Wilson-Fisher fixed points $(c_*^2,\frac{\epsilon_\tau}{2(n+8)},0,0,0)$, where $c_*^2$ is arbitrary and can be set to unity by independent redefinitions of the fermion and boson fields. We also have the GNY fixed points, for all $n=1,2,3$ and $N$ given by:
\begin{align}\label{CFP}
\left(1,\frac{4-n-N+\sqrt{D_C}}{4(n+8)(N+4-n)}\epsilon_\tau,\frac{\epsilon_\tau}{(N+4-n)},0,0\right),
\end{align}
where $D_C=N^2+2(5n+28)N+(4-n)^2$, in agreement with earlier studies~\cite{Zerf2017}. The fixed-point couplings are positive for all $N>0$. Since $c_*^2=1$ and $f(1)=\frac{1}{2}$ (Fig.~\ref{fig:f}), Eq.~(\ref{z}) implies that the CFPs are Lorentz invariant ($z=1$), and are in fact conformally invariant.

We next turn to DFPs, for which $\Delta_*$ and/or $v_*$ are nonzero. To be physical, all fixed points must obey the following conditions~\cite{Weinrib1983}:
\begin{align}\label{physicality}
c^2_*>0,\hspace{5mm}
\lambda^2_*\geq 0,\hspace{5mm}
h^2_*\geq 0,\hspace{5mm}
v_*\geq 0,\hspace{5mm}
\Delta_*+v_*\geq 0.
\end{align}
At fermionic DFPs with $h_*^2>0$, the condition $\beta_{c^2}=0$ together with Eq.~(\ref{physicality}) further implies that $c_*^2>1$. From Eq.~(\ref{bc2}), we find that at a fermionic fixed point,
\begin{align}\label{c2>1}
N(c_*^2-1)+2nc_*^2\left(f(c_*^2)-\half\right)=\frac{2(\Delta_*+v_*)c_*^2}{h_*^2}.
\end{align}
Equation~(\ref{physicality}) implies that the right-hand side of this equation is positive. From Fig.~\ref{fig:f} and Eq.~(\ref{f}), we see that $f(c_*^2)>\half$ only if $c_*^2>1$, and $f(c_*^2)<\half$ only if $c_*^2<1$. Thus for the left-hand side of Eq.~(\ref{c2>1}) to be positive also we must have $c_*^2>1$. (At a clean fermionic fixed point, the left-hand side must vanish, which can only happen for $c_*^2=1$.)

\subsubsection{Fixed points with short-range correlated disorder}
\label{sec:SDFPs}

We first focus on DFPs with $\Delta_* \neq 0$ and $v_*=0$, which we term short-range disordered fixed points (SDFPs). From Eq.~(\ref{bh2}) we find that $h_*^2=0$ or $h_*^2=\epsilon_\tau/(N+4-n)$. When the fixed-point value of the Yukawa coupling is zero, we reproduce the results of Refs.~\cite{Boyanovsky1982,boyanovsky1983,Lawrie1984} for the purely bosonic $O(n)$ vector model with random-mass disorder. For $n=1$, there is an accidental degeneracy in the system of equations $\beta_{\lambda^2}=0, \beta_{\Delta}=0$. The degeneracy is lifted at two-loop order, giving rise to a DFP with $\lambda_*^2,\Delta_*\sim\c{O}(\sqrt{\epsilon_\tau})$, for a finite ratio $\epsilon/\epsilon_\tau$~\cite{Boyanovsky1982}.

Our focus, however, is on fermionic DFPs with nonzero $h_*^2$. We find two fermionic SDFPs for $n=2,3$:
\begin{align}\label{SDFP1}
\left(c_{*\text{1,2}}^2,\frac{N+8-2n \pm \sqrt{D_S}}{8(n-1)(N+4-n)}\epsilon_\tau,\frac{\epsilon_\tau}{N+4-n},\frac{(n+2)(N \pm \sqrt{D_S})+2(4-n)^2}{16(n-1)(N+4-n)}\epsilon_\tau,0\right),
\end{align}
where $D_S=N^2-4(5n-8)N+4(4-n)^2$, which we denote by SDFP1 (with $+\sqrt{D_S}$, $c_*^2=c_{*1}^2$) and SDFP2 (with $-\sqrt{D_S}$, $c_*^2=c_{*2}^2$). The chiral XY case ($n=2$) was discussed in our earlier work~\cite{Yerzhakov2018}: the fixed-point couplings $\lambda_*^2$, $h_*^2$, and $\Delta_*$ are nonnegative, and thus physical, for all $N\geq 1$. At $N=1$, SDFP2 merges with the clean GNY fixed point (\ref{CFP}), while SDFP1 runs off to infinity as it is impossible to satisfy $\beta_{c^2}=0$. (Note that for $n=2$, SDFP1,2 here correspond to DFP1,2 in Ref.~\cite{Yerzhakov2018} for $N<4$ and to DFP2,1 for $N>4$.) In the chiral Heisenberg case ($n=3$), the discriminant $D_S \geq 0$ for $N \geq N_D \approx 27.856$, and the SDFPs (\ref{SDFP1}) are physical only for $N>N_D$.

In the chiral Ising case ($n=1$), as previously mentioned the RG equations for $\lambda^2$ and $\Delta$ become degenerate for zero Yukawa coupling, and we find only one solution at order $\c{O}(\epsilon, \epsilon_\tau)$ for $h_*^2 \neq 0$:
\begin{align}\label{SDFP_m=0}
\left( c_*^2, \frac{N \epsilon_\tau}{(N+3)(N+6)},\frac{\epsilon_\tau}{N+3},\frac{3(N-6)\epsilon_\tau}{4(N+3)(N+6)},0 \right).
\end{align}
This SDFP is physical for $N \geq 6$, and merges with the clean GNY fixed point at $N=6$. There is in principle the possibility of an additional SDFP at two-loop order with $\lambda_*^2,\Delta_*\sim\c{O}(\sqrt{\epsilon_\tau})$, as in the bosonic case, and $h_*^2\sim\c{O}(\epsilon_\tau)$. We show in Appendix~\ref{app:NoPoint} that this cannot happen, because it is impossible to satisfy the equation $\beta_{c^2}=0$. We also note that this excludes the possibility of a physical SDFP for the $N=1/2$ chiral Ising GNY model, which in the clean limit flows to a conformal field theory with emergent supersymmetry~\cite{sonoda2011,fei2016}, the $\c{N}=1$ Wess-Zumino model. (This theory describes the time-reversal symmetry-breaking transition among the gapless Majorana surface states of a three-dimensional topological superfluid, e.g., $^{3}$He-$B$~\cite{Grover2014}.)

For the fermionic SDFPs found in Eqs.~(\ref{SDFP1}-\ref{SDFP_m=0}) above, despite the fact that the equation $\beta_{c^2}=0$ is nonlinear in $c^2$, one can show analytically that it admits a unique solution $c_*^2>1$, except for $N=1$ in the XY GNY model. The actual fixed-point values of $c^2$ are obtained by solving the equation numerically, and together with $h_*^2$ determine via Eq.~(\ref{z}) the dynamic critical exponent $z$ at those fixed points (see Sec.~\ref{sec:exponents}, Fig.~\ref{fig:SDFPs_z}).

\subsubsection{Fixed points with long-range correlated disorder}
\label{sec:LDFPs}

We now turn to DFPs with $v_* \neq 0$, which we dub long-range disordered fixed points (LDFPs). For vanishing $h_*^2$, the purely bosonic random-mass $O(n)$ vector model for $n > 1$ was studied in the triple epsilon expansion in Ref.~\cite{DeCesare1994}, where LDFPs were found. For $n=1$, long-range correlated disorder lifts the previously mentioned degeneracy in the system of fixed-point equations. For nonzero $h_*^2=\epsilon_\tau/(N+4-n)$, we find two fermionic LDFPs in all three GNY universality classes, $n=1,2,3$: 
\begin{align}
\lambda^2_{*1,2} &=\frac{3(N+4-n) \delta - (5N+4-n) \epsilon_\tau  \pm \sqrt{D_L} }{4(5n+4) (N+4-n)},\label{L1}\\
(\Delta_*+v_*)_{1,2} &=\frac{ -2(n-1) (N+4-n) \delta + \bigl[ (5n-2)N-9+(n-1)^2\bigr] \epsilon_\tau  \pm (2+n) \sqrt{D_L} }{4 (5n+4) (N+4-n)},\label{L2}\\
v_{*1,2} &= \left(1+\frac{ 4 (\Delta_*+v_*)_{1,2}}{2\epsilon_\tau-\delta}\right) (\Delta_*+v_*)_{1,2},\label{L3}
\end{align}
where $D_L= \left[( 5N + 4 - n )\epsilon_\tau - 3 (N+4-n)\delta  \right]^2 - 8(5n+4)N\epsilon_\tau^2 $. The discriminant $D_L$ is nonnegative, and thus the fixed-point couplings real, for
either:
\begin{align}\label{deltaD}
\delta \geq \delta_{D}\equiv\frac{ (5N+4-n)+\sqrt{8(5n+4)N} }{3(N+4-n)}\epsilon_\tau,
\end{align}
or:
\begin{align}\label{deltaD2}
\delta \leq \delta_D'\equiv\frac{ (5N+4-n)-\sqrt{8(5n+4)N} }{3(N+4-n)}\epsilon_\tau.
\end{align}

In addition to being real, the fixed-point couplings (\ref{L1}-\ref{L3}) must obey the conditions (\ref{physicality}). By contrast with the SDFPs (\ref{SDFP1}-\ref{SDFP_m=0}), which are physical above a certain critical value of $N$ that is independent of $\delta$, the LDFPs are physical only in complicated regions of the $N$-$\delta$ plane that possess several disconnected components and/or curved boundaries. Since the fixed-point couplings (\ref{L1}-\ref{L3}) do not depend explicitly on $c_*^2$, we first assume a physical solution for $c_*^2$ exists, and discuss how the remaining conditions delimit those nontrivial regions.

\begin{itemize}
\item $\underline{\lambda^2_* \geq 0}$: This condition is satisfied for all $n=1,2,3$ for both LDFPs provided that $\delta \geq \delta_{D}$. Since $\delta_D>\delta_D'$ for all $N>0$, LDFPs in the region $\delta\leq\delta_D'$ of Eq.~(\ref{deltaD2}) are never physical.
\item $\underline{\Delta_*+v_* \geq 0}$: For LDFP1, i.e., Eqs.~(\ref{L1}-\ref{L3}) with $+\sqrt{D_L}$, the condition is satisfied for different regions of the $N$-$\delta$ plane depending on $n$:
\begin{align}
n=1:\,&\delta \in \begin{cases} [0, \delta_2] \cup [\delta_1, 4\epsilon_\tau], &   N \leq N_2, \\ [0, \delta_D'] \cup [\delta_D, 4\epsilon_\tau], & N > N_2; \end{cases}\\
n=2,3:\,& \delta \in [0,\delta_D'] \cup \begin{cases} [\delta_1, 4\epsilon_\tau], &   N \leq N_2, \\  [\delta_D, 4\epsilon_\tau], & N > N_2.
\end{cases}
\end{align}
\iffalse
\begin{align}
\delta \in [0,\delta_D'] \cup \begin{cases} [\delta_1, 4\epsilon_\tau], &   N \leq N_2, \\  [\delta_D, 4\epsilon_\tau], & N > N_2. \end{cases}
\end{align}
\fi
For LDFP2, i.e., Eqs.~(\ref{L1}-\ref{L3}) with $-\sqrt{D_L}$, we have:
\begin{align}
n=1:\,&\delta \in  \cup  \begin{cases} 
\varnothing, & N<N_2, \\
[\delta_{2}, \delta_D'] \cup [\delta_{D},\delta_1], & N \geq N_2;  \end{cases}  \\
n=2,3:\,& \delta \in \begin{cases} [\delta_2, \delta_D'], & N \leq N_2,  \\  [\delta_2, \delta_D'] \cup [\delta_D, \delta_1], & N \geq N_2. \end{cases} 
\end{align}
Here,
\begin{align}
\delta_{1}&\equiv\frac{[(n+14)N+9-(n-1)^2] + (n+2) \sqrt{D_C} }{(n+8)(N+4-n)} \epsilon_\tau,\label{delta1}\\
\delta_{2}&\equiv\frac{[(n+14)N+9-(n-1)^2] - (n+2) \sqrt{D_C} }{(n+8)(N+4-n)} \epsilon_\tau,
\end{align}
and $N_2$ is the value of $N$, which depends on $n$, at which $\delta_1=\delta_D$. For $N<N'<N_2$, $\delta_D'<0$, in which case $[0,\delta_D']$ denotes the empty set. We use the same notational convention whenever the left limit of the interval is greater than the right one.
\item $\underline{v_* \geq 0}$: For LDFP1, we have the following constraints depending on the value of $n$:
\begin{align}
n=1:\,&\delta \in \begin{cases} [0, \delta_2] \cup [\delta_1, 2\epsilon_\tau), &   N \leq N_2, \\ [0, \delta_D'] \cup [\delta_D, 2\epsilon_\tau), & N > N_2; \end{cases}  \\ 
n=2:\,&\delta \in [0,\delta_D'] \cup [\delta_D, 2\epsilon_\tau) \cup 
\begin{cases}
 [\delta_1, \delta_4) \cup [\delta_3, 4\epsilon_\tau],    1 \leq N < N_2, \\ 
 [\delta_D, \delta_4] \cup [\delta_3, 4\epsilon_\tau],      N_2 \leq N \leq N_3, \\
 [\delta_3, 4\epsilon_\tau],       N > N_3;
\end{cases} \\
n=3:\,&\delta \in [0,\delta_D'] \cup [\delta_D, 2\epsilon_\tau) \cup 
\begin{cases}
 [\delta_1, 4\epsilon_\tau],    1 \leq N < N_2, \\ 
 [\delta_D, 4\epsilon_\tau],    N_2 \leq N < N_D, \\ 
 [\delta_D, \delta_4] \cup [\delta_3, 4\epsilon_\tau],      N_D \leq N \leq N_3, \\
 [\delta_3, 4\epsilon_\tau],       N > N_3.
\end{cases}
\end{align}
For LDFP2, we have:
\begin{align}
n=1:\,&\delta \in [\delta_5, \max(2\epsilon_\tau,\delta_1)] \cup \begin{cases} \varnothing, 1 \leq N < N_2, \\  [\delta_{2},\delta_D'] \cup [\delta_{D},\min(\delta_1, 2\epsilon_\tau)], N \geq N_2; \end{cases} \\
n=2,3:\,&\delta \in [\delta_2, \delta_D'] \cup [\delta_D, 2\epsilon_\tau) \cup 
\begin{cases}
 \varnothing,    1 \leq N < N_2, \\ 
 [\delta_D, \delta_1],      N_2 \leq N < N_3, \\
 [\delta_4, \delta_1],       N \geq N_3.
\end{cases}\label{EqLDFP2}
\end{align}
\iffalse
\begin{align}
n=1:\,&\delta \in \begin{cases}
 \textcolor{red}{[\delta_5, 2\epsilon_\tau],    1 \leq N < N_2,}\\ 
 [\delta_2, \delta_D'] \cup  [\delta_D, \delta_1] \cup [\delta_5, 2\epsilon_\tau],      N_2 \leq N < 6,\\
 [\delta_2, \delta_D'] \cup [\delta_D, 2\epsilon_\tau),       6 \leq N \leq N'_1,\\
 [\delta_2, \delta_D'], N'_1<N<N'_2,\\ 
 [\delta_2, \delta_D'] \cup [\delta_D, 2\epsilon_\tau], N \geq N'_2, 
\end{cases}\cup  
\begin{cases}
 \varnothing,    1 \leq N < 6, \\ 
 [\delta_5, \delta_1], N \geq 6;
\end{cases}  \\
n=2,3:\,&\delta \in [\delta_2, \delta_D'] \cup [\delta_D, 2\epsilon_\tau) \cup 
\begin{cases}
 \varnothing,    1 \leq N < N_2, \\ 
 [\delta_D, \delta_1],      N_2 \leq N < N_3, \\
 [\delta_4, \delta_1],       \textcolor{red}{N \geq N_3.}
\end{cases}\label{EqLDFP2}
\end{align}
\fi
We further define
\begin{align}
\delta_{3}&\equiv\frac{3[N+6+(n-1)(3N+6-2n)] + (n+2)\sqrt{D_S} }{4(n-1)(N+4-n)} \epsilon_\tau,\label{delta3}\\
\delta_{4}&\equiv\frac{3[N+6+(n-1)(3N+6-2n)] - (n+2)\sqrt{D_S} }{4(n-1)(N+4-n)} \epsilon_\tau,\label{delta4}\\
\delta_5&\equiv\frac{2N^2+21N+18}{(N+3)(N+6)}\epsilon_\tau,\label{delta03}
\end{align}
and $N_3$ is the $n$-dependent value of $N$ at which $\delta_D=\delta_4$.
\end{itemize}
For a given GNY symmetry class $n$, the intersection of all those conditions defines regions in the $N$-$\delta$ plane in which the various fixed points discussed are physical, and over which fixed-point properties are plotted throughout the paper.

We now return to the question of whether a physical solution $c_*^2$ to the nonlinear equation $\beta_{c^2}=0$ exists for the LDFPs (\ref{L1}-\ref{L3}). We solve this equation numerically. For $n=1$ and $n=3$, we find a unique solution everywhere in the physical regions of the $N$-$\delta$ plane. For $n=2$, we likewise find a unique physical solution in the physical regions, but for LDFP1 computations become increasingly difficult upon approach to the point $N=1$, $\delta=4$, where $c_*^2$ grows rapidly. Since exactly at this point LDFP1 coincides with SDFP2, and SDFP2 does not admit a solution to $\beta_{c^2}=0$ for $N=1$~\cite{Yerzhakov2018}, we conjecture that $c_*^2$ gradually runs off to infinity as the point $N=1$, $\delta=4$ is approached. Summarizing, we thus find that for all three GNY symmetry classes, a unique solution $c_*^2>1$ exists for the LDFPs (\ref{L1}-\ref{L3}) everywhere inside the physical regions (\ref{physicality}) of the $N$-$\delta$ plane. As mentioned previously, $h_*^2$ and $c_*^2$ together determine the dynamic critical exponent $z$ at those fixed points (Sec.~\ref{sec:exponents}, Figs.~\ref{fig:LDFPs_z_m=0}-\ref{fig:LDFPs_z_m=2}).

\subsection{Linear stability analysis}
\label{sec:stability}

We now investigate the stability properties of the physical fixed points. All bosonic fixed points (i.e., with $h_*^2=0$) are unstable with respect to the $h^2$ direction. Additionally, for all models, the Gaussian fixed points are unstable with respect to all other directions, and the Wilson-Fisher fixed points are unstable with respect to both short-range and long-range correlated disorder. The stability properties of the bosonic DFPs in the absence of Yukawa coupling have been discussed previously in Refs.~\cite{Boyanovsky1982,boyanovsky1983,Lawrie1984,DeCesare1994}.

At all fermionic fixed points (i.e., with $h_*^2 \neq 0$), the $h^2$ direction is irrelevant. Additionally, we find that $\partial\beta_{c^2}/\partial c^2$ is positive at all such fixed points. Since $\beta_{c^2}$ is the only beta function in which $c^2$ appears, this means $c^2$ is also an irrelevant direction. We can thus exclude $h^2$ and $c^2$ from RG flow considerations and investigate stability within the three-dimensional subspace with fixed $h_*^2$ and $c_*^2$ of the full five-dimensional space of couplings. We compute the eigenvalues $y$ of the stability matrix $M_{gg'}\equiv-\partial\beta_g/\partial g'$, $g,g'\in\{\lambda^2,\Delta,v\}$, defined such that $y>0$ ($y<0$) corresponds to a relevant (irrelevant) direction.

\subsubsection{Stability of the clean fixed point}
\label{sec:StabCFP}

We first focus on the clean GNY fixed point (\ref{CFP}), which for the rest of the paper we refer to as the CFP. The RG eigenvalues at the CFP are:
\begin{align}\label{y4CFP}
y_1=-\frac{\sqrt{D_C}}{N+4-n}\epsilon_\tau, \hspace{5mm}
y_2=\frac{(n+2)N+(n+14)(4-n)-(n+2)\sqrt{D_C}}{(n+8)(N+4-n)}\epsilon_\tau, \hspace{5mm}
y_3=\delta-\delta_1,
\end{align}
and are associated with eigenvectors with nonzero projections along the $\lambda^2$, $\Delta$, and $v$ directions, respectively. The eigenvalue $y_1$ is negative and thus irrelevant for all $n$ and $N$. For the flow of short-range correlated disorder ($y_2$), we discuss the three GNY symmetry classes in turn.
\begin{itemize}
\item $n=1$: Disorder is irrelevant for $N>6$. At $N=6$, the CFP merges with the SDFP (\ref{SDFP_m=0}), and disorder becomes marginally relevant. For $N<6$ (including $N=1/2$), the SDFP becomes unphysical, and disorder becomes relevant at the CFP.
\item $n=2$: This case was studied in Ref.~\cite{Yerzhakov2018}. Disorder is irrelevant for $N>1$. At $N=1$, SDFP2 [see Eq.~(\ref{SDFP1})] merges with the CFP and disorder becomes marginally relevant.
\item $n=3$: Disorder is irrelevant for all $N>\frac{2}{15}\approx 0.133$.
\end{itemize}

Finally, long-range correlated disorder ($y_3$) is irrelevant for $\delta$ less than $\delta_1$, which is defined in Eq.~(\ref{delta1}). At generic points along the curve $\delta=\delta_1$ in the $N$-$\delta$ plane, one of the LDFPs merges with the CFP, and long-range correlated disorder crosses marginality. At the special point $N=N_2$ along this curve, the two LDFPs (\ref{L1}-\ref{L3}) coincide with one another (and with the CFP).

\subsubsection{Stability of short-range disordered fixed points}
\label{sec:StabSDFPs}

We now consider the SDFPs of Sec.~\ref{sec:SDFPs}. We begin with the unique SDFP (\ref{SDFP_m=0}) in the chiral Ising class ($n=1$), which is physical only for $N\geq 6$. Long-range correlated disorder is irrelevant at this fixed point provided that $\delta$ is less than $\delta_5$, which is defined in Eq.~(\ref{delta03}). Along the curve $\delta=\delta_5$ in the $N$-$\delta$ plane, the SDFP merges with LDFP2. However, one of the two other eigenvalues is always relevant for $N>6$, thus the SDFP is a multicritical point with at least one relevant direction on the critical hypersurface.

\begin{figure}[!t]
\includegraphics[width=0.99\columnwidth]{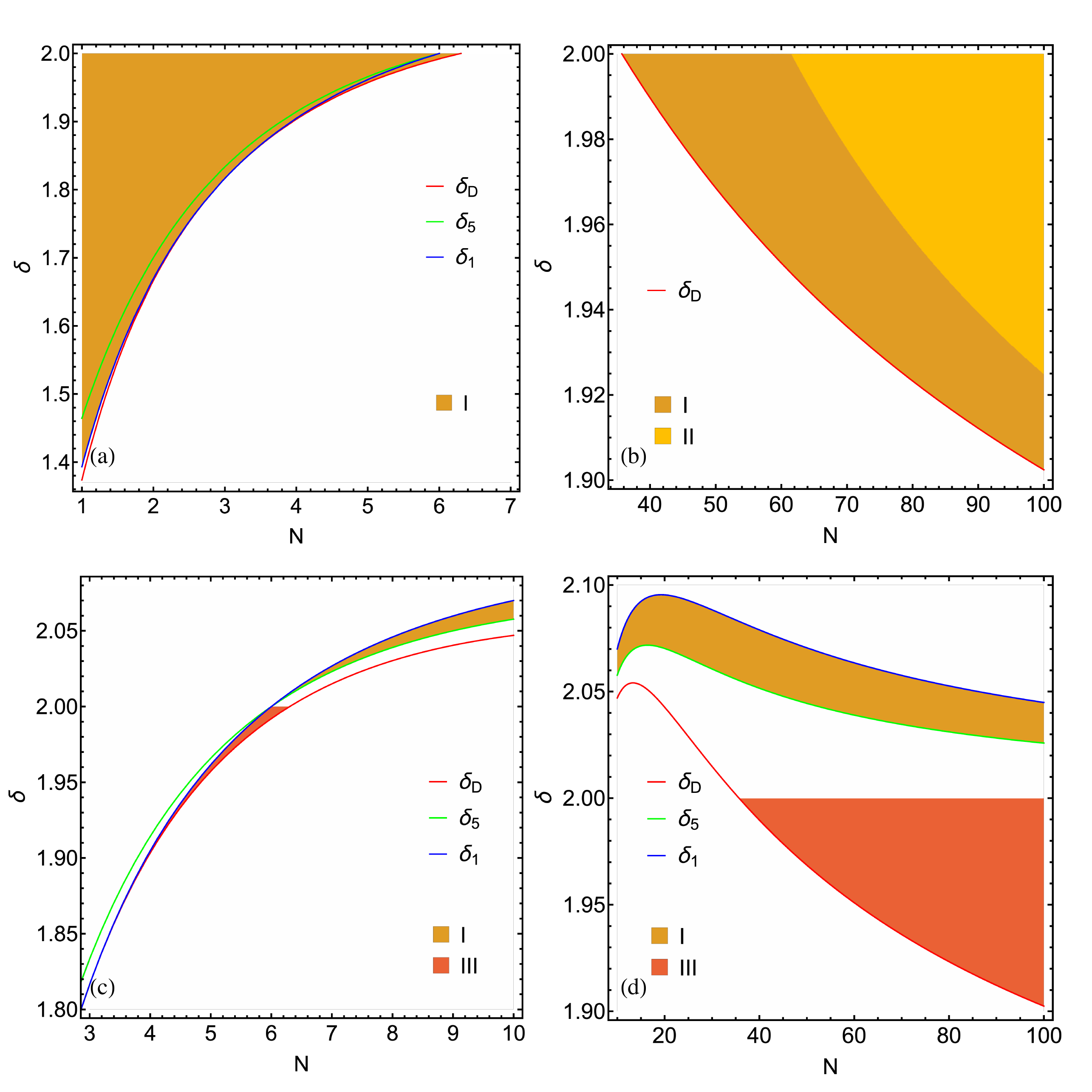}
\caption{Stability in the subspace $(\lambda^2, \Delta, v)$ of couplings of (a,b) LDFP1 and (c,d) LDFP2 in the chiral Ising GNY model ($n=1$), as a function of $N$ and $\delta$. I: one relevant eigenvalue; II: one relevant eigenvalue, two complex-conjugate irrelevant eigenvalues; III: two relevant eigenvalues.}
\label{fig:EVs_m=0}
\end{figure}

\begin{figure}[!t]
\includegraphics[width=0.99\columnwidth]{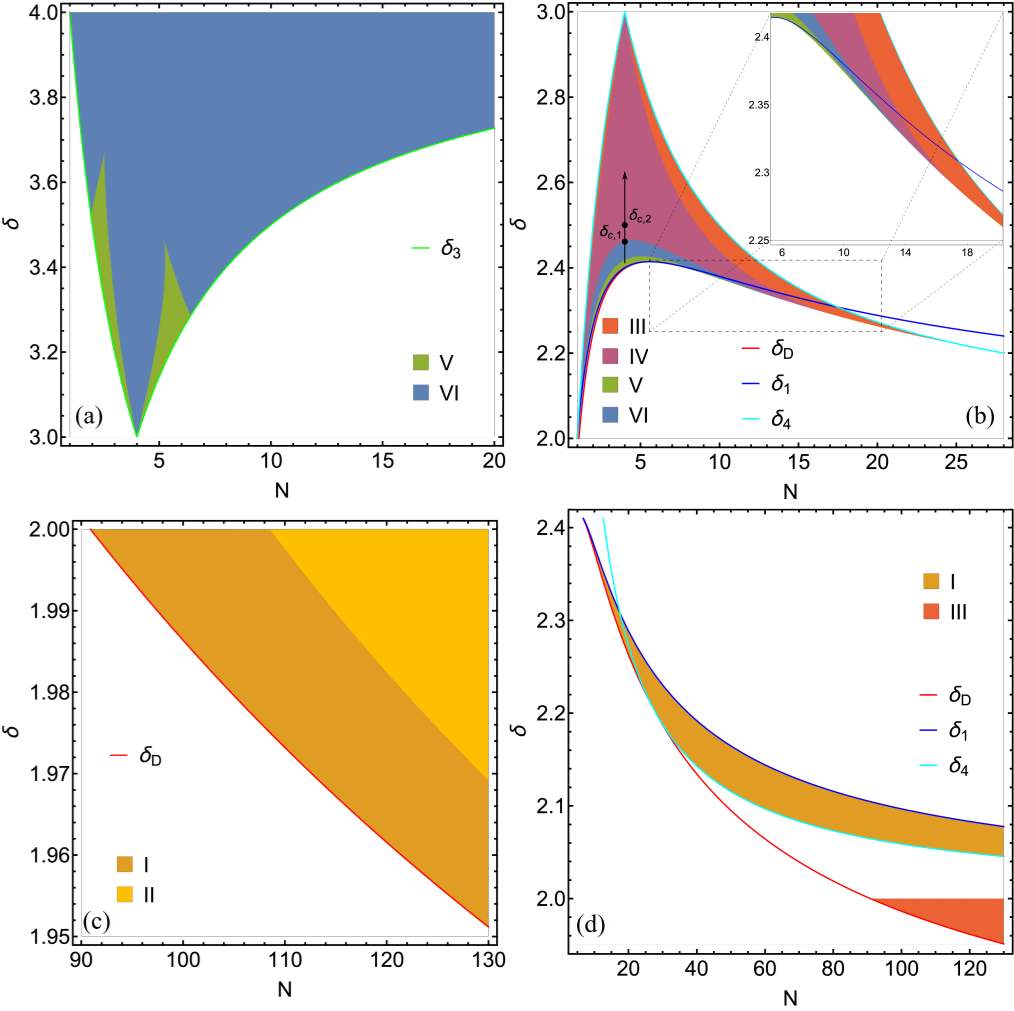}
\caption{Stability in the subspace $(\lambda^2, \Delta, v)$ of couplings of (a,b) LDFP1 and (c,d) LDFP2 in the chiral XY GNY model ($n=2$), as a function of $N$ and $\delta$. Regions I-III are defined as in Fig.~\ref{fig:EVs_m=0}. IV: two complex-conjugate relevant eigenvalues; V: no relevant eigenvalues; VI: no relevant eigenvalues, two complex-conjugate irrelevant eigenvalues.}
\label{fig:EVs_m=1}
\end{figure}

\begin{figure}[!t]
\includegraphics[width=0.99\columnwidth]{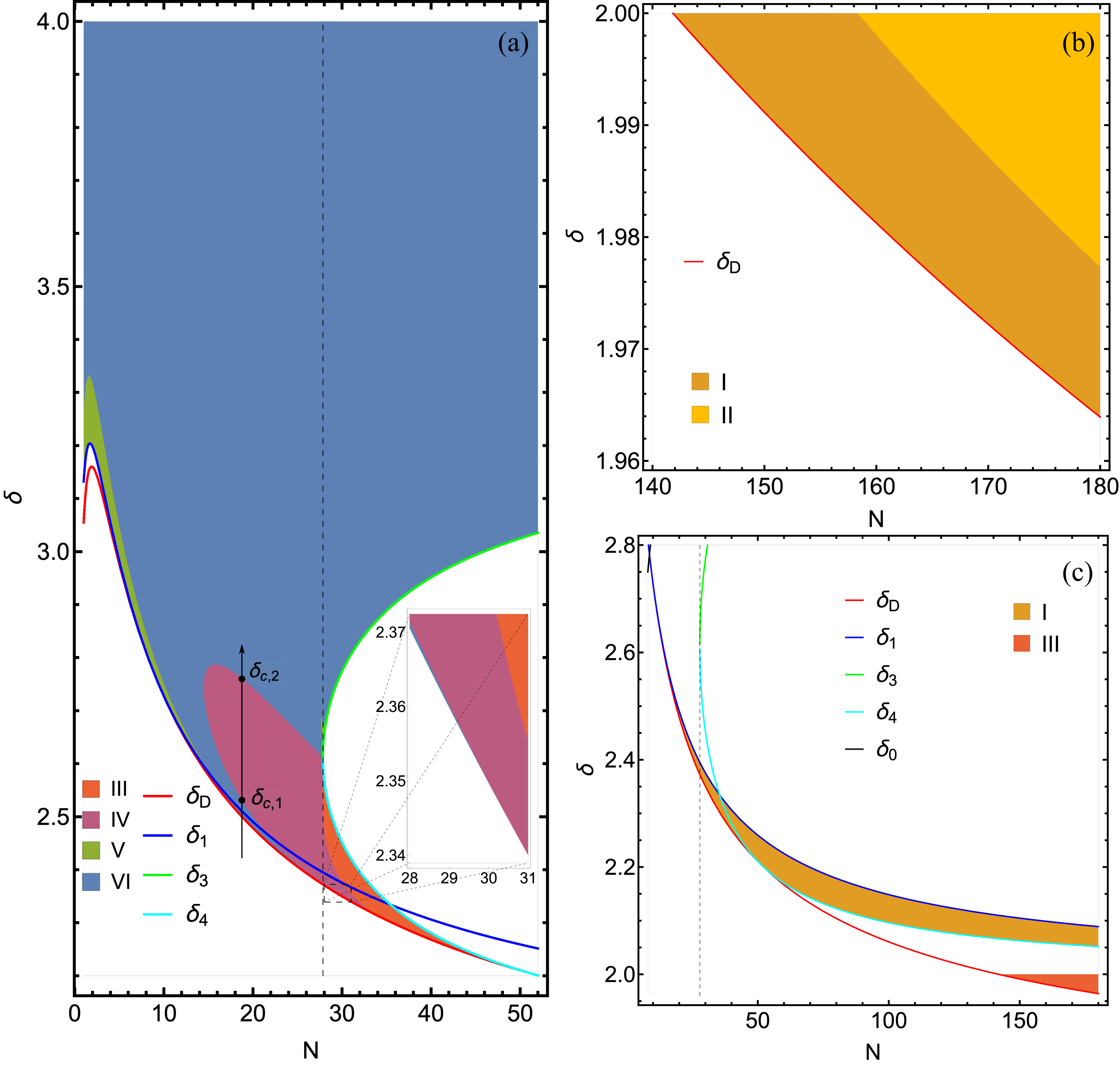}
\caption{Stability in the subspace $(\lambda^2, \Delta, v)$ of couplings of (a,b) LDFP1 and (c) LDFP2 in the chiral Heisenberg GNY model ($n=3$), as a function of $N$ and $\delta$. Regions are labeled as in Fig.~\ref{fig:EVs_m=1}.}
\label{fig:EVs_m=2}
\end{figure}

The chiral XY ($n=2$) and Heisenberg ($n=3$) classes admit two fermionic SDFPs, Eq.~(\ref{SDFP1}). Similarly to the chiral Ising case, long-range correlated disorder is irrelevant at SDFP1 (SDFP2) provided that $\delta<\delta_3$ ($\delta<\delta_4$), with $\delta_3,\delta_4$ defined in Eqs.~(\ref{delta3}-\ref{delta4}). The curves $\delta=\delta_3$ and $\delta=\delta_4$ correspond to the merger of the corresponding SDFP with one of the LDFPs. When $\delta_3=\delta_4$, the discriminant $D_S$ vanishes, and the two SDFPs merge with one another. This happens at a critical value of $N$ which in the XY case is $N=4$, and in the Heisenberg case is $N=N_D\approx 27.856$. Besides long-range correlated disorder, the other two directions are irrelevant at SDFP1, thus it is a genuine critical point for $\delta<\delta_{3}$. By contrast, one of those two directions is relevant at SDFP2, thus the latter is a multicritical point.

For the chiral XY and Heisenberg models, and for sufficiently large $N$, the two irrelevant eigenvalues at SDFP1 with eigenvectors in the $\lambda^2$-$\Delta$ plane form a complex conjugate pair. SDFP1 is then a fixed point of focus type, with spiraling flows near the fixed point. In the XY case, this happens for $N>\frac{32}{5}=6.4$, while for the Heisenberg case, this happens for $N>28.087$. Critical properties in this case are subject to oscillatory corrections to scaling~\cite{Khmelnitskii1978,Yerzhakov2018}.

\subsubsection{Stability of long-range disordered fixed points}

We finally turn to the stability of the LDFPs of Sec.~\ref{sec:LDFPs}. The eigenvalues of the stability matrix depend on $N$ and $\delta$ in a complicated way, and we compute them numerically. In Figs.~\ref{fig:EVs_m=0}-\ref{fig:EVs_m=2}, we characterize the stability of the two LDFPs in terms of their number of relevant/irrelevant eigenvalues, for each GNY symmetry class. Eigenvalues are real unless otherwise specified; since the stability matrix is real, complex eigenvalues necessarily appear in complex-conjugate pairs, and imply focus-type behavior as discussed above. For all three GNY symmetry classes, the two LDFPs merge along the curve $\delta=\delta_D$ in the $N$-$\delta$ plane, where the discriminant $D_L$ vanishes. In the Ising case (Fig.~\ref{fig:EVs_m=0}), both LDFPs have at least one relevant eigenvalue on the critical hypersurface and are thus multicritical points (for $N=1/2$, only LDFP1 is physical, for $\delta_1\approx 1.143<\delta<2$). In the XY and Heisenberg cases (Figs.~\ref{fig:EVs_m=1}-\ref{fig:EVs_m=2}), LDFP1 exists in regions (V and VI) in the $N$-$\delta$ plane with no relevant eigenvalues, and is thus a {\it bona fide} critical point in those regions. LDFP2 is always multicritical.

\subsection{Critical exponents}
\label{sec:exponents}

Universal critical exponents at the newly found fermionic DFPs can be computed from Eqs.~(\ref{nuinv}-\ref{gammaphi}) using the fixed-point couplings found in Sec.~\ref{sec:SDFPs} and Sec.~\ref{sec:LDFPs}. At the present one-loop order, the fermion $\eta_\psi$ and boson $\eta_\phi$ anomalous dimensions depend only on $h_*^2$, which is the same at all fermionic fixed points. Thus their values at the DFPs are the same as those for the clean chiral GNY universality classes~\cite{Zerf2017}: $\eta_\psi=n\epsilon_\tau/[2(N+4-n)]$ and $\eta_\phi=N\epsilon_\tau/(N+4-n)$. At higher loop order the anomalous dimensions are expected to differ at the different fermionic fixed points.

\begin{figure}[!t]
\includegraphics[width=0.7\columnwidth]{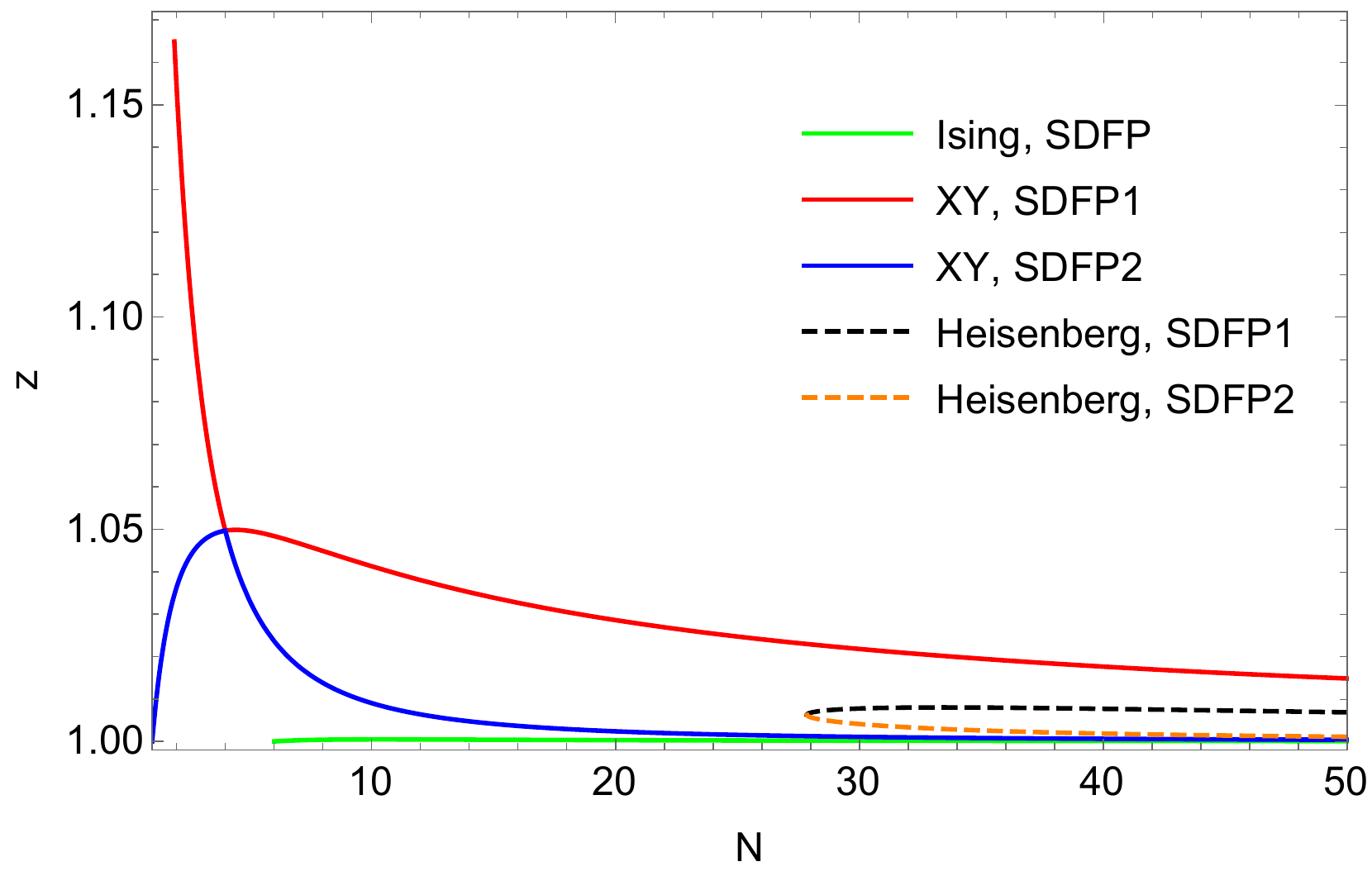}
\caption{Dynamic critical exponent $z$ at SDFPs for all three chiral GNY symmetry classes, as a function of $N$.}
\label{fig:SDFPs_z}
\end{figure}

\begin{figure}[!t]
\includegraphics[width=0.99\columnwidth]{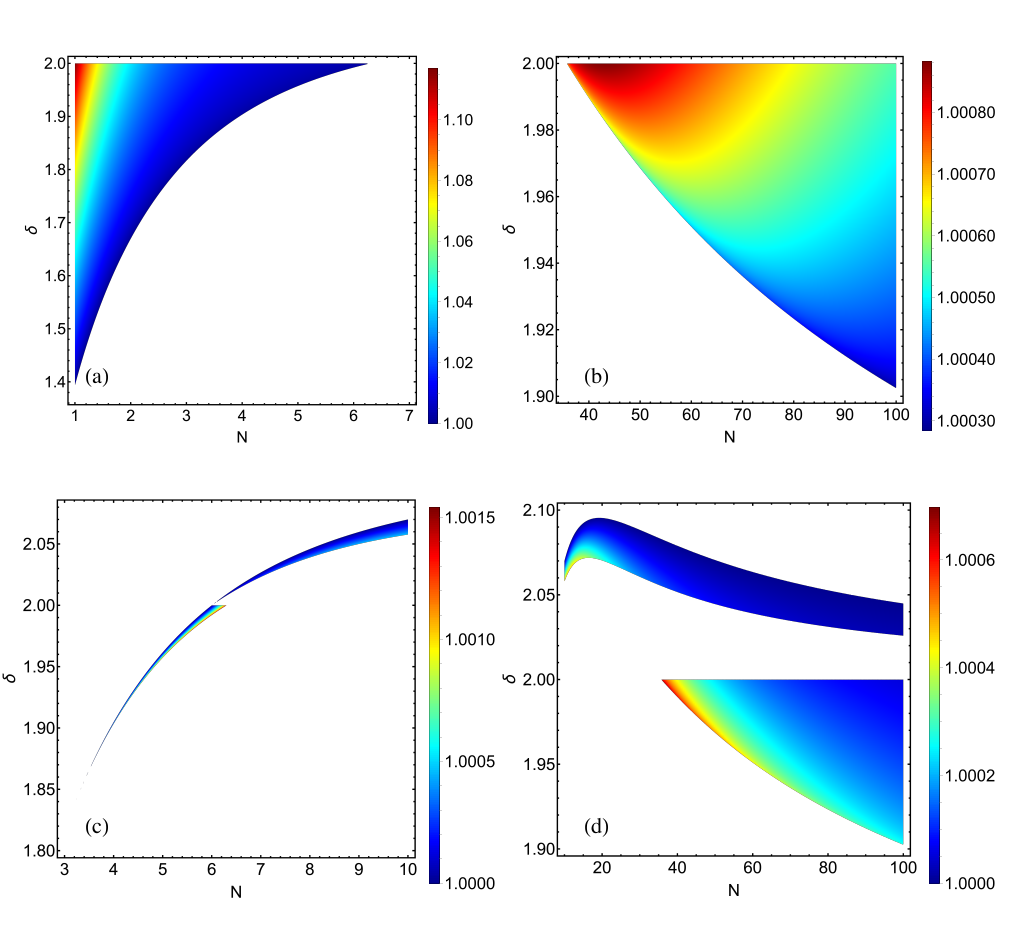}
\caption{Dynamic critical exponent $z$ in the chiral Ising GNY model ($n=1$) at (a,b) LDFP1 and (c,d) LDFP2, as a function of $N$ and $\delta$.}
\label{fig:LDFPs_z_m=0}
\end{figure}

\begin{figure}[!t]
\includegraphics[width=0.99\columnwidth]{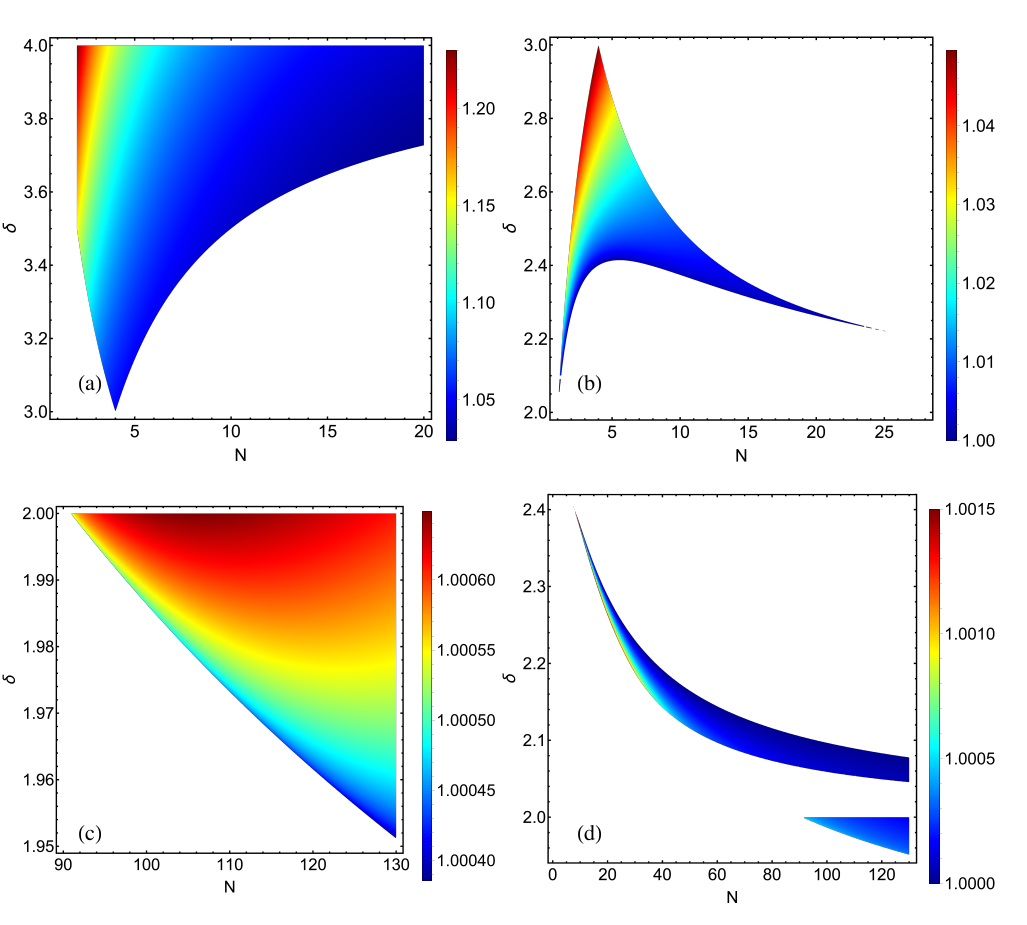}
\caption{Dynamic critical exponent $z$ in the chiral XY GNY model ($n=2$) at (a,b) LDFP1 and (c,d) LDFP2, as a function of $N$ and $\delta$.}
\label{fig:LDFPs_z_m=1}
\end{figure}

\begin{figure}[!t]
\includegraphics[width=0.99\columnwidth]{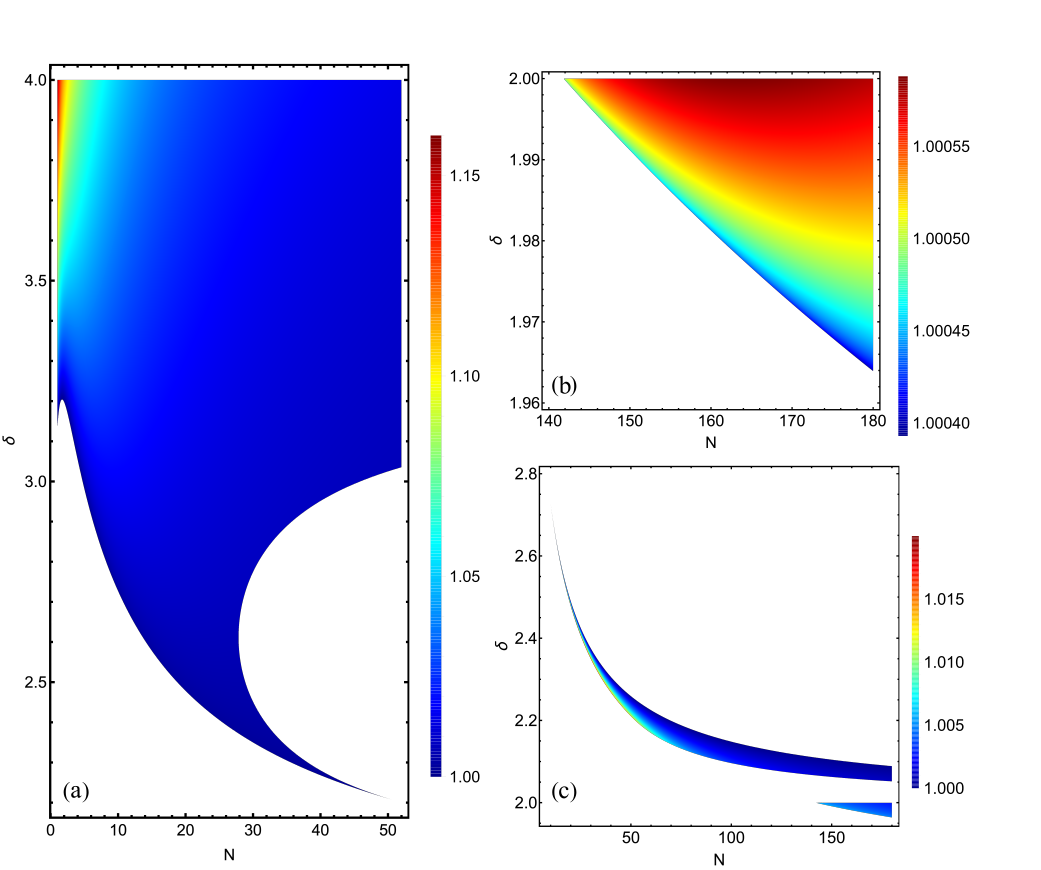}
\caption{Dynamic critical exponent $z$ in the chiral Heisenberg GNY model ($n=3$) at (a,b) LDFP1 and (c) LDFP2, as a function of $N$ and $\delta$.}
\label{fig:LDFPs_z_m=2}
\end{figure}

Using Eq.~(\ref{z}), the dynamic critical exponent $z$ at the fermionic DFPs is given by
\begin{align}
z=1+\left(f(c_*^2)-\half\right)\frac{n\epsilon_\tau}{N+4-n},
\end{align}
and thus depends on the fixed-point velocity parameter $c_*^2$. The latter is a universal function of $N$ and $\delta$ for a given DFP but must be computed numerically; we plot the resulting value of $z$ extrapolated to 2+1 dimensions ($\epsilon_\tau\rightarrow 1$) in Fig.~\ref{fig:SDFPs_z} for the SDFPs and in Figs.~\ref{fig:LDFPs_z_m=0}-\ref{fig:LDFPs_z_m=2} for the LDFPs. Since $c_*^2>1$, and thus $f(c_*^2)>\half$, at all fermionic DFPs (see Sec.~\ref{sec:FixedPoints}), such DFPs necessarily have $z>1$. This is in agreement with the general expectation that weak disorder increases $z$~\cite{herbut2001}; Refs.~\cite{narovlansky2018,aharony2018} also derive the leading-order result $z-1\propto\Delta_*>0$ at SDFPs obtained by perturbing a conformally invariant QCP with weak short-range correlated disorder. Here we find $z>1$ at LDFPs as well.

The inverse correlation length exponent $\nu^{-1}$, determined from Eq.~(\ref{nuinv}), is the RG eigenvalue associated with the relevant direction $r$ which tunes across the symmetry-breaking transition. For a \emph{bona fide} critical point, $\nu$ controls the divergence of the correlation length $\xi$ at the transition $r=0$ via $\xi\sim r^{-\nu}$. For multicritical points with additional relevant directions $g_1,g_2,\ldots$ on the critical hypersurface with real, positive eigenvalues $y_1,y_2,\ldots$, the correlation length behaves near the transition as $\xi(r,g_1,g_2,\ldots)=r^{-\nu}\widetilde{\xi}(g_1/r^{\nu y_1},g_2/r^{\nu y_2},\ldots)$, where $\widetilde{\xi}(x_1,x_2,\ldots)$ is a universal scaling function~\cite{Goldenfeld}. Complex-conjugate eigenvalues produce a scaling function with oscillatory behavior. At all LDFPs in all three GNY symmetry classes, we find $\nu^{-1}=2-\half\delta$, which alternatively can be written as $\nu=2/\alpha$, with $\alpha=4-\delta$ the exponent controlling long-range disorder correlations in Eq.~(\ref{DisCorr}). This superuniversal behavior was also found at long-range correlated bosonic DFPs and explained by Weinrib and Halperin~\cite{Weinrib1983}. Consider a LDFP with correlation length exponent $\nu(\alpha)$ in a system with disorder of the type (\ref{DisCorr}). If one further perturbs this fixed point with disorder correlated according to $|\b{x}-\b{x}'|^{-\alpha_+}$ such that $\alpha_+>\alpha$, the original asymptotic critical behavior should remain the same, as we expect it is controlled by the longest-range part of the disorder. Conversely, if the perturbation is of the form $|\b{x}-\b{x}'|^{-\alpha_-}$ with $\alpha_-<\alpha$, this falls off more slowly than the original disorder, and the original critical behavior should be unstable. Assuming $\alpha,\alpha_\pm<d$ and applying the modified Harris criterion for long-range correlated disorder, we find $\nu(\alpha)>2/\alpha_+$ and $\nu(\alpha)<2/\alpha_-$, for all $\alpha_-<\alpha<\alpha_+$. Choosing $\alpha_\pm=\alpha\pm\varepsilon$ and taking the limit $\varepsilon\rightarrow 0^+$, we obtain $\nu(\alpha)=2/\alpha$.

\begin{figure}[!t]
\includegraphics[width=0.9\columnwidth]{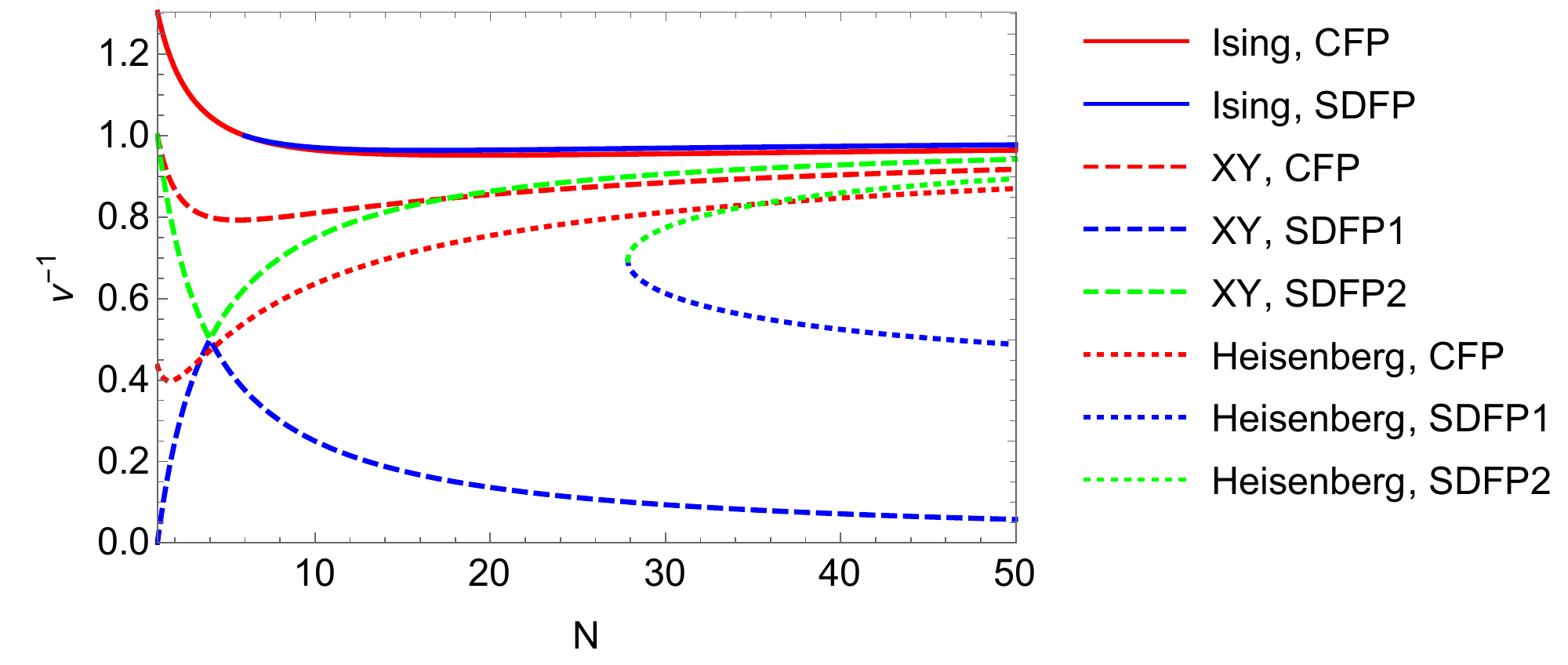}
\caption{Inverse correlation length exponent $\nu^{-1}$ for the CFP and SDFPs in all three chiral GNY symmetry classes, as a function of $N$.}
\label{fig:Inverse_nu_Allm}
\end{figure}

The exponent $\nu$ for the SDFPs can likewise be calculated directly from Eq.~(\ref{nuinv}), and we obtain $\nu^{-1}=2-\half\delta_5$ for the chiral Ising SDFP, with $\delta_5$ defined in Eq.~(\ref{delta03}). In light of the result above for $\nu^{-1}$ at LDFPs, this is consistent with the fact that the $n=1$ SDFP coalesces with one of the LDFPs at $\delta=\delta_5$. Similarly, for both the chiral XY and Heisenberg models we find that SDFP1 has $\nu^{-1}=2-\half\delta_3$ and SDFP2 has $\nu^{-1}=2-\half\delta_4$, with $\delta_{3,4}$ defined in Eqs.~(\ref{delta3}-\ref{delta4}). As previously mentioned, the curves $\delta=\delta_3$ ($\delta=\delta_4$) correspond to the merger of SDFP1 (SDFP2) with a LDFP. We plot $\nu^{-1}$ at SDFPs for all three GNY models in Fig.~\ref{fig:Inverse_nu_Allm}, including $\nu^{-1}$ at the clean GNY critical point for comparison.

\section{RG flows and bifurcations}
\label{sec:RGflows}

Having discussed RG fixed points and their local properties (stability and critical exponents), we now discuss global properties of the RG flow: bifurcations of the flow as the control parameters $N,\delta$ are varied (Secs.~\ref{sec:bifurc1} and \ref{sec:bifurc2}), and examples of global phase diagrams for fixed $N,\delta$ (Sec.~\ref{sec:phasediagram}). Although the original space of couplings $(c^2,\lambda^2,h^2,\Delta,v)$ is five-dimensional, as already mentioned the $c^2$ and $h^2$ directions are irrelevant at fermionic fixed points, which are the only stable ones. For practical purposes the RG flows thus live in the three-dimensional space $(\lambda^2,\Delta,v)$, with $c^2$ and $h^2$ assuming their fixed-point values. Since in the chiral Ising case all physical fixed points are multicritical, and for the sake of simplicity, we restrict our attention to the chiral XY and Heisenberg symmetry classes, which exhibit the most interesting phenomena.

\subsection{Transcritical and saddle-node bifurcations}
\label{sec:bifurc1}

We have already mentioned a number of instances in which two fixed points collide as $N$ or $\delta$ are varied. We observe two distinct kinds of bifurcations associated with a collision of two fixed points: the transcritical bifurcation and the saddle-node bifurcation.

\begin{figure}[!t]
\includegraphics[width=0.9\columnwidth]{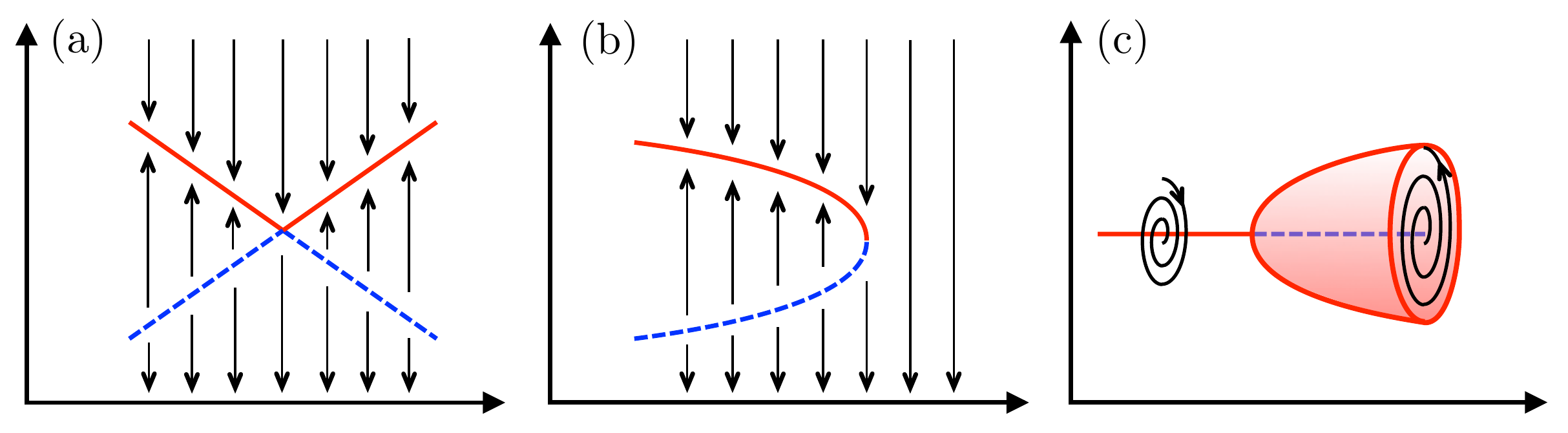}
\caption{Schematic bifurcation diagrams for (a) the transcritical bifurcation, (b) the saddle-node bifurcation, and (c) the supercritical Hopf bifurcation. The horizontal axis represents a direction in the $N$-$\delta$ plane, and the vertical axis, the space of running couplings (critical hypersurface). Solid red symbolizes an RG attractor, dashed blue a repellor, and schematic RG trajectories are shown in black.}
\label{fig:bifurcation}
\end{figure}

The transcritical bifurcation [Fig.~\ref{fig:bifurcation}(a)] is a bifurcation at which a stable fixed point and an unstable fixed point pass through each other, exchanging their stability properties, but without annihilating~\cite{gukov2017}. An example of this bifurcation is the merging of the two chiral XY SDFPs (\ref{SDFP1}) as $N$ is varied through $N=4$. (There is ``exchange'' of fixed points provided we track individual fixed points on smooth trajectories, as opposed to their arbitrary definition as SDFP1 and SDFP2 in Eq.~(\ref{SDFP1}).) Unlike the saddle-node bifurcation discussed below, the two fixed points remain real before and after the bifurcation. At the transcritical bifurcation, the beta function (and associated RG flow) is not only marginal, but its derivative with respect to the control parameter, here $N$, must vanish as well. Other examples of this bifurcation include the collision of SDFPs with the CFP (at $N=1$ for the chiral XY SDFP2), of LDFPs with the CFP (along the curve $\delta=\delta_1$ in the $N$-$\delta$ plane), or of SDFPs with LDFPs (curves $\delta=\delta_3$ and $\delta=\delta_4$). At these latter bifurcations, one of the DFPs becomes unphysical, by either $\Delta_*$, $v_*$, or $\Delta_*+v_*$ going through zero and becoming negative. However, since the other fixed point remains physical and thus real, this unphysical fixed point necessarily remains real also (for another RG example of this scenario, see Ref.~\cite{boyack2018}). Thus the bifurcation is distinct from the saddle-node bifurcation, which we now discuss.

The saddle-node bifurcation [Fig.~\ref{fig:bifurcation}(b)] is a bifurcation at which a stable fixed point and an unstable fixed point merge, leading to marginal behavior as above, but subsequently disappear into the complex plane. This typically happens for a pair of fixed points with critical couplings $g_{*\pm}\propto A\pm\sqrt{D}$, such that the discriminant $D$ continuously goes through zero at the bifurcation and then becomes negative. Both pairs SDFP1,2 and LDFP1,2 are of this type. The two chiral Heisenberg SDFPs, with discriminant $D=D_S(n=3)$, annihilate with decreasing $N$ at $N\approx 27.856$. (For the chiral XY GNY model, $D=D_S(n=2)$ touches zero at $N=4$ but remains positive elsewhere, which gives the transcritical bifurcation at $N=4$.) Likewise, the two LDFPs in both the XY and Heisenberg cases annihilate on the curve $\delta=\delta_D$ in the $N$-$\delta$ plane, where the discriminant $D=D_L$ vanishes. Since $\delta_D$ in Eq.~(\ref{deltaD}) is a nonmonotonic function of $N$, for fixed $\delta$ this fixed-point annihilation can occur for either increasing or decreasing $N$.

\begin{figure}[!t]
\includegraphics[width=\columnwidth]{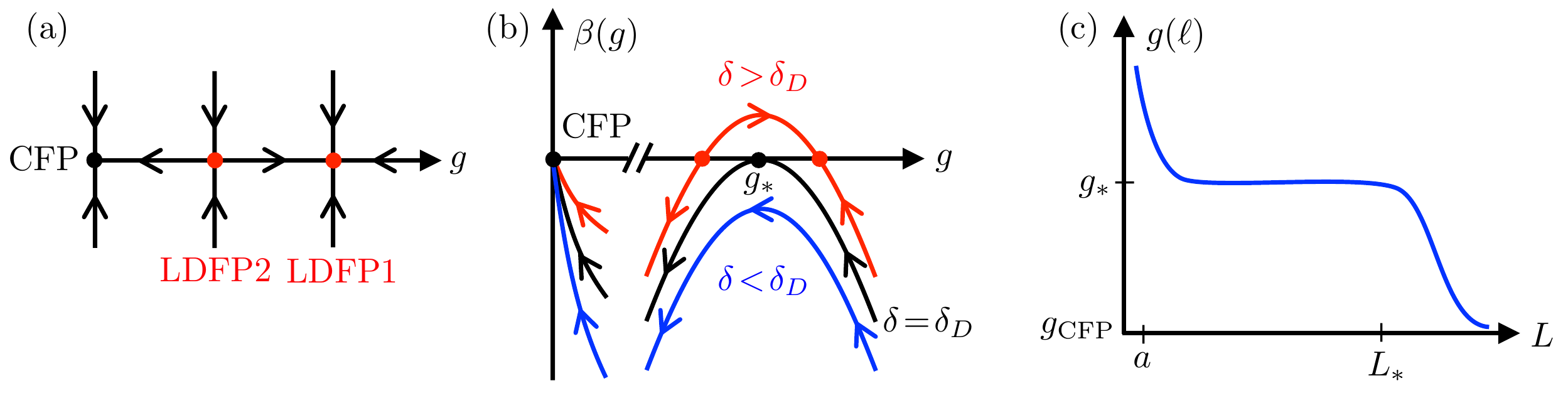}
\caption{Phenomenology of the saddle-node bifurcation at $\delta=\delta_D$. (a) Curvilinear coordinate $g$ along RG trajectories for $\delta>\delta_D$; (b) Wilsonian beta function near the bifurcation; (c) crossover from disordered quasi-critical behavior to clean critical behavior for $\delta$ slightly below $\delta_D$.}
\label{fig:quasiqc}
\end{figure}

The saddle-node bifurcation is accompanied by the characteristic phenomenology of walking RG or quasi-critical behavior~\cite{Kaplan2009}; we now explain how this manifests itself in the current problem. Focusing on the example above of the annihilation of LDFPs in the chiral XY and Heisenberg GNY models, we first consider a situation where $\delta$ is slightly above $\delta_D$. Small regions in the $N$-$\delta$ plane exist such that both LDFPs are physical, with LDFP1 a stable sink-type fixed point (region V) and LDFP2 a multicritical point with one relevant direction (region I). LDFP2 is only physical provided $\delta<\delta_1$ [see Eq.~(\ref{EqLDFP2})], which implies that the CFP is stable (Sec.~\ref{sec:StabCFP}). For this type of region, numerical studies of the RG flow show that RG trajectories with initial conditions near LDFP2 end up at either LDFP1 or the CFP. We thus consider a curvilinear coordinate system such that one of these coordinates, $g$, passes through all three fixed points [Fig.~\ref{fig:quasiqc}(a)]. In this section only, we define the infrared (Wilsonian) beta function $\beta(g)\equiv dg/d\ell$, where $\ell$ grows towards the infrared. Denoting by $g_*$ the common fixed-point coupling of LDFP1 and LDFP2 at the bifurcation $\delta=\delta_D$, we assume that for $\delta$ near $\delta_D$ and $g$ near $g_*$, $\beta(g)$ can be well approximated by a quadratic function, $\beta(g)\approx A(\delta)+B(\delta)(g-g_*)+C(\delta)(g-g_*)^2$. Since $\beta(g_*)=\partial\beta(g_*)/\partial g=0$ and $\partial^2\beta(g_*)/\partial g^2<0$ at $\delta=\delta_D$, we have $A(\delta_D)=B(\delta_D)=0$ and $C(\delta_D)\equiv -\kappa<0$. For $\delta=\delta_D+\varepsilon$ with $\varepsilon$ small, $\beta(g)$ should have two real zeros that approach $g_*$ as $\varepsilon\rightarrow 0^+$. Expanding $A(\delta)$, $B(\delta)$, and $C(\delta)$ in powers of $\varepsilon$, we find at leading order a pair of zeros of the form $g_*\pm\sqrt{b\varepsilon/\kappa}$ with $b\equiv A'(\delta_D)$, which are real provided that $b>0$, and form a complex-conjugate pair when $\varepsilon<0$ ($\delta<\delta_D$). The beta function thus approximately assumes the form $\beta(g)\approx b(\delta-\delta_D)-\kappa(g-g_*)^2$, illustrated in Fig.~\ref{fig:quasiqc}(b), and considered in Ref.~\cite{Kaplan2009}.

We now take $\delta=\delta_D-\varepsilon$ with $\varepsilon>0$ small, and consider an RG trajectory with initial coupling $g_\text{UV}>g_*$ and ``flow velocity'' $\beta(g_\text{UV})$, which is generically not small. As $g$ approaches $g_*$ from above, the flow velocity decreases considerably (i.e., the running coupling ``walks''), since $\beta(g_*)\approx -b\varepsilon$ is small. This walking behavior persists until $g_*-g$ becomes on the order of $\sqrt{b\varepsilon/\kappa}$, after which the coupling starts ``running'' again. This determines a characteristic RG time $\Delta\ell$ insensitive to the initial condition $g_\text{UV}$ of the flow. Approximating $\beta(g)\approx\beta(g_*)\approx -b\varepsilon$ as constant during the walk, we have $\beta(g_*)\approx\Delta g/\Delta\ell\sim\sqrt{b\varepsilon/\kappa}/\Delta \ell$, and thus $\Delta\ell\sim 1/\sqrt{\kappa b\varepsilon}$. Alternatively, we may integrate the equation $dg/d\ell=\beta(g)$ from $g_\text{UV}$ at $\ell_\text{UV}$ to $g_\text{IR}<g_*$ at $\ell_\text{IR}$. Under the condition $|g_\text{UV,IR}-g_*|\gg\sqrt{b\varepsilon/\kappa}$, the result of this integration is insensitive to the precise values of $g_\text{UV}$ and $g_\text{IR}$, and we obtain $\Delta\ell\equiv\ell_\text{IR}-\ell_\text{UV}=\pi/\sqrt{\kappa b\varepsilon}$. In turn, this RG time determines a characteristic infrared length scale $L_*=L_\text{IR}=L_\text{UV}e^{\Delta\ell}$, where we can take $L_\text{UV}\sim a$ to be on the order of a microscopic lattice constant $a$. We obtain:
\begin{align}
L_*\sim a\exp\left(\pi/\sqrt{\kappa b(\delta_D-\delta)}\right),
\end{align}
as $\delta$ approaches $\delta_D$ from below. The exponential inverse-square-root divergence, reminiscent of the divergence of the correlation length at the Kosterlitz-Thouless transition~\cite{kosterlitz1974}, is characteristic of the saddle-node bifurcation~\cite{Kaplan2009}. The existence of this exponentially large length scale $L_*\gg a$ allows for a crossover between two distinct physical regimes [Fig.~\ref{fig:quasiqc}(c)]. On intermediate length scales $a\ll L\ll L_*$, RG trajectories dwell for an extended period of RG time near $g=g_*$, and we have quasi-critical behavior controlled by a complex pair of LDFPs with real part near $g_*$. This quasi-critical regime is characterized by approximate power-law scaling and drifting (i.e., scale-dependent) exponents~\cite{gorbenko2018b}. On the largest length scales $a\ll L_*\ll L$, the transition is controlled by the true infrared fixed point, the CFP, with genuine scale invariance.

\subsection{Supercritical Hopf bifurcation and limit-cycle fermionic quantum criticality}
\label{sec:bifurc2}

The third type of bifurcation we observe is the supercritical Hopf bifurcation [Fig.~\ref{fig:bifurcation}(c)]. This bifurcation occurs as one passes from region VI (blue region) to region IV (purple region) in both the chiral XY [Fig.~\ref{fig:EVs_m=1}(b)] and Heisenberg [Fig.~\ref{fig:EVs_m=2}(a)] models. For instance, one can consider keeping $N$ fixed and tuning $\delta$ (black arrow in those figures). In region VI ($\delta<\delta_{c,1}$), LDFP1 is a stable-focus fixed point with two complex-conjugate irrelevant eigenvalues, i.e., complex-conjugate eigenvalues with a negative real part [solid red line on left part of Fig.~\ref{fig:bifurcation}(c)]. At the bifurcation ($\delta=\delta_{c,1}$), the real part of those eigenvalues goes through zero and becomes positive for $\delta>\delta_{c,1}$. LDFP1 thus loses its stability and becomes an unstable-focus fixed point [dashed blue line on the right part of Fig.~\ref{fig:bifurcation}(c)]. At the same type, a stable limit cycle is born [solid red line on the right part of Fig.~\ref{fig:bifurcation}(c)], towards which the spiraling RG trajectories coming out of LDFP1 asymptote, and which controls the critical behavior up to a second threshold value $\delta_{c,2}$ to be discussed shortly. (Trajectories outside the limit cycle also spiral and asymptote to it.) 

To our knowledge, this is the first instance in the context of quantum phase transitions where the supercritical Hopf bifurcation~\cite{Marsden1976} appears. After Ref.~\cite{hartnoll2016}, which studied a holographic model of a critical scalar field perturbed by disorder, our result is the second example of quantum phase transition governed by a stable limit cycle; to our knowledge, it is the first example for fermionic systems. The subcritical Hopf bifurcation~\cite{Marsden1976}, where an unstable-focus fixed point becomes stable by giving birth to an unstable limit cycle, has been reported previously in RG studies of classical disordered systems~\cite{Weinrib1983,Athorne1985,*Athorne1986}. The general phenomenology of critical behavior controlled by a stable limit cycle was explored in Ref.~\cite{Veytsman1993}. For a stable-focus critical point, spiraling trajectories manifest themselves as oscillatory corrections to scaling~\cite{Khmelnitskii1978,Yerzhakov2018}. By contrast, for a transition governed by a stable limit cycle, thermodynamic quantities exhibit log-periodic scaling behavior at leading order, i.e., discrete scale invariance. For instance, we show in Appendix~\ref{app:LimitCycleScaling} that the order parameter susceptibility $\chi$ obeys the approximate scaling form:
\begin{align}\label{ChiLimitCycle}
\chi\sim|r|^{-\gamma_\text{LC}}\left[1+\gamma_\text{LC}\c{F}\left(\nu_\text{LC}\ln\left(\frac{r_0}{r}\right)\right)\right],
\end{align}
where $\c{F}$ is a periodic function. Here $\nu_\text{LC}$ and $\gamma_\text{LC}=(2-\eta_\phi)\nu_\text{LC}$ are effective correlation-length and susceptibility exponents for the limit cycle, $r$ is the tuning parameter for the transition, and $r_0$ is a nonuniversal constant.

As $\delta$ is further increased past $\delta_{c,1}$, the limit cycle eventually disappears at a second critical value $\delta_{c,2}$, but in different ways for the chiral XY and Heisenberg GNY models. In the Heisenberg case, the Hopf bifurcation of Fig.~\ref{fig:bifurcation}(c) occurs again but in reverse: the limit cycle shrinks to a point, which becomes the stable-focus LDFP1 of region VI. In the XY case, our numerical studies suggest that at least for some values of $N$, the limit cycle is destroyed at $\delta=\delta_{c,2}$ (still within region IV) by colliding with the CFP and SDFP2, which are both saddle points in this regime [see Fig.~\ref{fig:phasediagram}(c)]. This is a possible example of heteroclinic bifurcation~\cite{dingjun1997}, whose detailed study we reserve for future work.

\subsection{Schematic phase diagrams}
\label{sec:phasediagram}

From the knowledge of the stability properties of the various fixed points and limit cycles, and numerical investigation of the RG flow connecting those different critical manifolds, schematic phase diagrams can be constructed analogously to those in Ref.~\cite{Yerzhakov2018}. For given values of $N$ and $\delta$, we focus on the critical hypersurface ($r=0$) and ask how the universality class of the transition depends on the bare couplings in the Lagrangian, which determine the initial conditions for the infrared RG flow. We consider a scenario in which the interaction parameters $h$ and $\lambda$ are fixed, and vary the two types of disorder, $\Delta$ and $v$. Since the number of possibilities is very large, given the complexity of the stability/physicality regions, we focus on the two most interesting regions: those which contain the instances of limit-cycle quantum criticality discussed in the previous section.

\begin{figure}[!t]
\includegraphics[width=0.8\columnwidth]{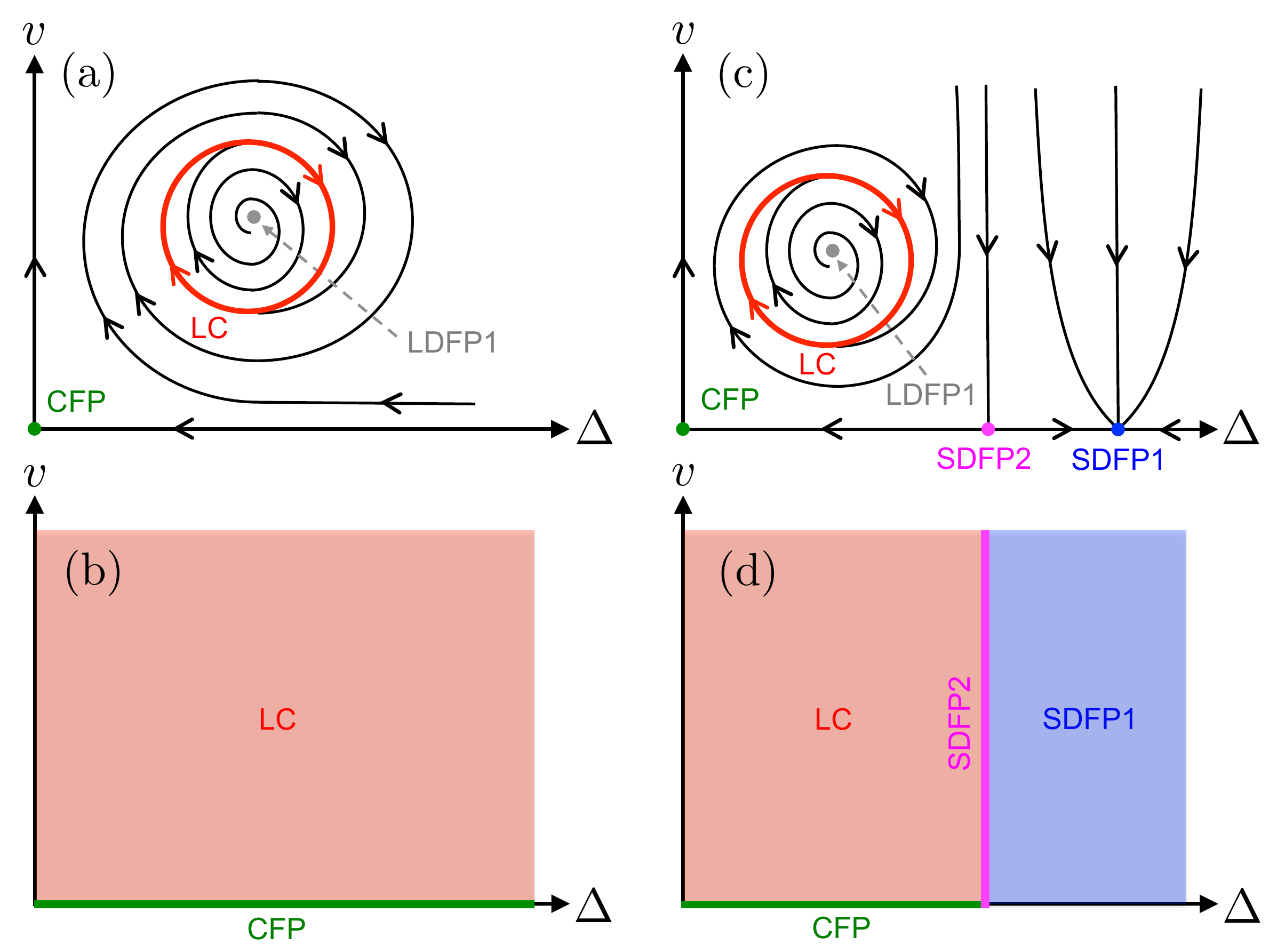}
\caption{Schematic RG flow and critical ($r=0$) phase diagrams for generic $N$ and $\delta_{c,1}<\delta<\delta_{c,2}$ in region IV (see Figs.~\ref{fig:EVs_m=1}-\ref{fig:EVs_m=2}), for (a,b) the chiral Heisenberg GNY model and (c,d) the chiral XY model. In the Heisenberg case, the transition is controlled by a stable limit cycle (LC) for generic bare values of the short-range correlated ($\Delta$) and long-range correlated ($v$) disorder strengths. In the XY case, the transition is controlled by the limit cycle for weak short-range disorder and by a disordered fixed point (SDFP1) for strong short-range disorder.}
\label{fig:phasediagram}
\end{figure}

We first focus on region IV in the chiral Heisenberg GNY model [see Fig.~\ref{fig:EVs_m=2}(a)]. For generic points in this region (e.g., for $\delta_{c,1}<\delta<\delta_{c,2}$), one has $\delta>\delta_D$ and $\delta>\delta_1$. Furthermore, we assume $N<N_D\approx 27.856$. From Sec.~\ref{sec:StabCFP}, we conclude that the CFP has two irrelevant directions in the $\lambda^2$-$\Delta$ plane, but that long-range correlated disorder $v$ is relevant, since $\delta>\delta_1$. SDFP1,2 are both unphysical, since $N<N_D$, and LDFP2 is unphysical as well. As seen in the previous section, LDFP1 is of unstable-focus type, with spiraling flow towards a stable limit cycle. The resulting RG flow is illustrated schematically in Fig.~\ref{fig:phasediagram}(a). Consequently, at least for sufficiently small bare values of the disorder, the transition is controlled by limit-cycle quantum criticality for generic disorder [Fig.~\ref{fig:phasediagram}(b)]. If long-range correlated disorder is turned off completely, the transition reverts back to the clean chiral Heisenberg GNY universality class.

We now turn to region IV in the chiral XY GNY model [see Fig.~\ref{fig:EVs_m=1}(b)], assuming $\delta_{c,1}<\delta<\delta_{c,2}$. As in the previous case, we generically have $\delta>\delta_D$, $\delta>\delta_1$, and also $\delta<\delta_4$. As in the Heisenberg case, the CFP has two irrelevant directions in the $\lambda^2$-$\Delta$ plane, but $v$ is relevant. There are now nontrivial SDFPs, whose stability was discussed in Sec.~\ref{sec:StabSDFPs}. For SDFP1, $\lambda^2$ and $\Delta$ are both irrelevant, and $v$ is irrelevant as well, since $\delta<\delta_4<\delta_3$. For SDFP2, $v$ is irrelevant since $\delta<\delta_4$, but there is one relevant direction with nonzero $\Delta$ projection. LDFP2 is unphysical, and LDFP1 is an unstable focus with flow towards a stable limit cycle. The resulting RG flow is schematized in Fig.~\ref{fig:phasediagram}(c), and the corresponding phase diagram in Fig.~\ref{fig:phasediagram}(d). For weak $\Delta$, the transition is governed by the limit cycle, but for sufficiently strong $\Delta$, the transition is controlled by a disordered fixed point, SDFP1. CFP and SDFP2 appear as multicritical points.

\section{Conclusion}
\label{sec:conclusion}

In summary, we have performed a comprehensive study of the three classes of chiral GNY models most relevant for symmetry-breaking quantum phase transitions in (2+1)D gapless Dirac matter---the chiral Ising, XY, and Heisenberg GNY models---in the presence of quenched short-range and long-range correlated random-mass disorder. Using a controlled triple epsilon expansion below the upper critical dimension for these models, we have found several disordered infrared fixed points characterized by finite short-range and/or long-range correlated randomness, and for which we computed critical exponents. The Boyanovsky-Cardy and quantum Weinrib-Halperin fixed points, while present, are destabilized by the Yukawa interaction in favor of new disordered fermionic QCPs, at which the strength of this interaction remains nonzero in the infrared. Besides local stability, using numerical and analytical approaches we analyzed bifurcations of the RG flow. We found instances of the familiar fixed-point annihilation scenario, which can here be tuned by a genuinely continuous variable---the exponent controlling the algebraic decay of disorder correlations---and with which is associated a parametrically large crossover length scale $L_*$ that separates a disordered quasi-critical regime ($L\ll L_*$) from a clean regime in the deep infrared ($L\gg L_*$). We also uncovered instances of the transcritical bifurcation, at which fixed points exchange their stability, and the more exotic supercritical Hopf bifurcation. The latter was accompanied by the emergence of a stable limit cycle on the critical hypersurface, thus producing the first instance of fermionic quantum criticality with discrete scale invariance.

Several avenues present themselves for future research. The relative paucity of disordered fixed points found in the chiral Ising class as compared to its continuous-symmetry counterparts, and in fact, the complete absence of {\it bona fide} critical points in this class, is in agreement with the conjecture by Motrunich {\it et al.}~\cite{motrunich2000} that all discrete symmetry-breaking transitions in (2+1)D disordered systems should fall in the infinite-randomness universality class. Since infinite-randomness fixed points are not accessible to perturbative RG methods, nonperturbative numerical studies of Ising transitions of interacting Dirac fermions with quenched randomness are desirable, e.g., using quantum Monte Carlo methods~\cite{ma2018} or, possibly, incorporating fermions into (2+1)-dimensional adaptations of the strong-disorder RG method~\cite{iyer2012}. In the presence of gapless Dirac fermions strongly coupled to bosonic order parameter fluctuations, rare-region effects~\cite{nandkishore2013,nandkishore2014}---which dominate the low-energy physics at infinite-randomness fixed points---may however lead to a different strong-disorder phenomenology than that found in local bosonic models~\cite{vojta2003}.

Besides the pure GNY universality classes, relevant to symmetry-breaking transitions in systems of itinerant Dirac electrons, our method of analysis may also provide a point of entry to study the effect of quenched disorder on more exotic transitions, such as those involving fractionalized phases. The algebraic or Dirac spin liquid~\cite{affleck1988,kim1999,rantner2001,*rantner2002,hermele2005,*hermele2007}, a quantum-disordered paramagnet with fractionalized spinon excitations, is described at low energies by (2+1)D quantum electrodynamics (QED$_3$) with $N=4$ flavors of two-component gapless Dirac fermions. The effect of quenched disorder on QED$_3$ itself was studied recently~\cite{Thomson2017,goswami2017,zhao2017,goldman2017,dey2020}; using the methods presented here, one could additionally study the effect of quenched disorder on quantum phase transitions out of the algebraic spin liquid~\footnote{As an experimental example of such transitions, Ref.~\cite{bordelon2019} reports the possible observation of a field-induced quantum phase transition between an algebraic spin liquid and a collinear magnetically ordered state in the triangular-lattice frustrated magnet NaYbO$_2$.}. Transitions towards conventional phases such as VBS states~\cite{boyack2019,zerf2020,janssen2020} or antiferromagnets~\cite{ghaemi2006,dupuis2019,zerf2019}, or transitions towards gapped chiral~\cite{janssen2017b,ihrig2018,zerf2018} or $\mathbb{Z}_2$ spin liquids~\cite{boyack2018}, are described by GNY theories in all three (Ising, XY, Heisenberg) symmetry classes, augmented by a coupling to fluctuating $U(1)$ gauge fields. The effect of random-mass disorder on the critical fixed points of such QED$_3$-GNY theories is an interesting topic for future research.

\acknowledgments

We thank D. A. Huse for a useful discussion. H.Y. was supported by Alberta Innovates and Alberta Advanced Education. J.M. was supported by NSERC Discovery Grants \#RGPIN-2014-4608, \#RGPIN-2020-06999, and \#RGPAS-2020-00064; the Canada Research Chair (CRC) Program; CIFAR; the Government of Alberta's Major Innovation Fund (MIF); the Tri-Agency New Frontiers in Research Fund (NFRF, Exploration Stream); and the Pacific Institute for the Mathematical Sciences (PIMS) Collaborative Research Group program.

\appendix

\numberwithin{equation}{section}
\numberwithin{figure}{section}

\section{Long-range correlated disorder contributions to the renormalization constants}
\label{app:Z}

In this Appendix we describe the computation of the renormalization constants $\delta Z_i\equiv Z_i-1$ from the one-loop diagrams in Fig.~\ref{fig:diagrams}. Diagrams in the clean limit or with short-range correlated disorder have been computed for the chiral XY GNY model in Ref.~\cite{Yerzhakov2018}, and are easily adapted to the entire family of GNY models. We only detail the computation of diagrams involving long-range correlated disorder.

\subsection{Boson two-point function}

Four diagrams contribute: Fig.~\ref{fig:diagrams}(a-d). Diagrams (a-b) appear in the pure GNY models, and have been well studied~\cite{Rosenstein1993,Zerf2017}. Diagram (c) appears in the purely bosonic random-mass $O(n)$ vector model~\cite{Boyanovsky1982,boyanovsky1983,Lawrie1984} and contributes to $\delta Z_3$ and $\delta Z_r$. Diagram (d) also contributes to $\delta Z_3$ and $\delta Z_r$, and we compute it here. Its contribution to the divergent part of the effective action is:
\begin{align}
\delta\Gamma_\text{div}^\text{(d)}&=-v \sum_a\int\frac{d^Dk}{(2\pi)^D} \b{\phi}_a(-k)\cdot\b{\phi}_a(k) \int\frac{d^d\b{p}}{(2\pi)^d}\frac{|\b{p}|^{\epsilon-\delta}}{c^2 k_0^2+(\b{k}+\b{p})^2+r\mu^2},
\end{align}
where $d^Dk=d^{\epsilon_\tau}k_0\,d^d\b{k}$, and we have discarded a term that vanishes in the replica $m \rightarrow 0$ limit. Since we anticipate a renormalization of both the time-derivative term~\cite{narovlansky2018,aharony2018} and the scalar mass term, the latter being necessary to compute the correlation length exponent, we must keep the ``mass squared'' $c^2k_0^2+r$ in the denominator. Such massive Feynman integrals can be evaluated using the Mellin-Barnes representation of hypergeometric functions~\cite{Boos1991,Dudka2015}. We have:
\begin{align}
I\equiv\int\frac{d^d\b{p}}{(2\pi)^d}\frac{|\b{p}|^{\epsilon-\delta}}{c^2 k_0^2+(\b{k}+\b{p})^2+r\mu^2}
&=(c^2k_0^2+r\mu^2)^{1-\delta/2}S_{4-\epsilon}\frac{\Gamma\left(-1+\frac{\delta}{2}\right)\Gamma\left(2-\frac{\delta}{2}\right)}{2\Gamma(1)}\nn\\
&\phantom{=}\times{}_2F_1\left(\frac{\delta-\epsilon}{2},-1+\frac{\delta}{2};2-\frac{\epsilon}{2};-\frac{\b{k}^2}{c^2k_0^2+r\mu^2}\right),
\end{align}
where ${}_2F_1(a,b;c;z)$ is the Gauss hypergeometric function, and $S_d=2/[(4\pi)^{d/2}\Gamma(d/2)]$. Taking the limit $\delta,\epsilon\rightarrow 0$, the hypergeometric function evaluates to a constant: ${}_2F_1(0,-1;2;z)=1$. The only divergent factor in this limit is $\Gamma(-1+\frac{\delta}{2})\rightarrow -2/\delta$, and we obtain:
\begin{align}
I=-\frac{2(c^2k_0^2+r\mu^2)}{(4\pi)^2\delta}.
\end{align}
After rescaling the couplings by $(4\pi)^2$, we thus obtain:
\begin{align}
\delta Z_3^\text{(d)}=\delta Z_r^\text{(d)}=-\frac{2v}{\delta}.
\end{align}

\subsection{Boson self-interaction}

Diagrams (g) and (h) are the same as in the pure GNY models, and diagram (i) only involves short-range correlated disorder. Diagram (j) contributes to the boson self-interaction vertex:
\begin{align}
\delta\Gamma_\text{div}^\text{(j)}&=  6 \lambda^2 v \sum_{a}
 \int\frac{d^Dk}{(2\pi)^D}\int\frac{d^Dk'}{(2\pi)^D}\int\frac{d^Dk''}{(2\pi)^D}
\,\phi_a^i(-k)\phi_a^j(-k')\phi_a^i(k'')\phi_a^j(k+k'-k'')\nn\\
&\phantom{=}\times\int\frac{d^d\b{p}}{(2\pi)^d}
\frac{|\b{p}|^{\epsilon-\delta}}
{\left[c^2(k_0+k_0'-k_0'')^2+(\b{k}+\b{k}'-\b{k}''+\b{p})^2+r\mu^2\right]
\left[c^2(k_0'')^2+(\b{k}''-\b{p})^2+r\mu^2\right]}.
\end{align}
Since we are looking for the correction to a local four-point vertex, we can set the external momenta $k,k',k''$ to zero in the integral over the loop momentum $\b{p}$. Using standard Euclidean integrals,
\begin{align}\label{EuclideanIntegral}
\int\frac{d^d\ell}{(2\pi)^d}\frac{\ell^m}{(\ell^2+\Delta^2)^n}=\frac{1}{(4\pi)^{d/2}}
\frac{\Gamma\left(\frac{d+m}{2}\right)\Gamma\left(n-\frac{d+m}{2}\right)}{\Gamma(d/2)\Gamma(n)}
\left(\frac{1}{\Delta^2}\right)^{n-(d+m)/2},
\end{align}
we then have, in the limit $\epsilon,\delta\rightarrow 0$,
\begin{align}\label{DimRegIntegral}
\int\frac{d^d\b{p}}{(2\pi)^d}\frac{|\b{p}|^{\epsilon-\delta}}{(\b{p}^2+r\mu^2)^2}=\frac{2}{(4\pi)^2\delta}.
\end{align}
Rescaling $v$ by $(4\pi)^2$, we obtain:
\begin{align}
\delta Z_5^\text{(j)}=-\frac{12v}{\delta}.
\end{align}

\subsection{Short-range correlated disorder strength}

Diagram (k) contributes to both the boson self-interaction vertex and the short-range correlated disorder vertex, and was computed before. Diagrams (l) and (m), which involve long-range correlated disorder, both contribute to the renormalization of the short-range disorder strength.

Diagrams of the type (l) give two distinct contributions, of the form:
\begin{align}\label{eqA9}
\delta \Gamma_\text{div}^\text{(l;1,2)}=-\Delta v\sum_{ab}
\int\frac{d^Dk}{(2\pi)^D}\int\frac{d^Dk'}{(2\pi)^D}\int\frac{d^d\b{k}''}{(2\pi)^d}
\,&\phi_a^i(-k)\phi_b^j(-k')\phi_a^i(\b{k}'',k_0)\phi_b^j(\b{k}+\b{k}'-\b{k}'',k_0')\nn\\
&\times\c{I}_{1,2}(k,k',\b{k}''),
\end{align}
where
\begin{align}
\c{I}_{1}(k,k',\b{k}'')&=2\int\frac{d^d\b{p}}{(2\pi)^d}
\frac{|\b{p}|^{\epsilon-\delta}}
{\left[c^2k_0^2+(\b{k}+\b{p})^2+r\mu^2\right]\left[c^2(k_0')^2+(\b{k}'-\b{p})^2+r\mu^2\right]},\label{eqA10}\\
\c{I}_{2}(k,k',\b{k}'')&=\int\frac{d^d\b{p}}{(2\pi)^d}
\frac{|\b{p}|^{\epsilon-\delta}}
{\left[c^2(k_0')^2+(\b{k}+\b{k}'-\b{k}''-\b{p})^2+r\mu^2\right]\left[c^2(k_0')^2+(\b{k}'-\b{p})^2+r\mu^2\right]}.
\end{align}
As in the previous section, we can set $k=k'=0,\,\b{k}''=0$ in those loop integrals, which then simply reduce to Eq.~(\ref{DimRegIntegral}). With $v$ rescaled by $(4\pi)^2$ as before, we then obtain:
\begin{align}
\delta Z_7^\text{(l)}=-\frac{12v}{\delta}.
\end{align}

Diagram (m) illustrates that long-range correlated disorder perturbatively generates short-range correlated disorder. We obtain:
\begin{align}
\delta\Gamma_\text{div}^\text{(m)}=-v^2\sum_{ab}
\int\frac{d^Dk}{(2\pi)^D}&\int\frac{d^Dk'}{(2\pi)^D}\int\frac{d^d\b{k}''}{(2\pi)^d}
\,\phi_a^i(-k)\phi_b^j(-k')\phi_a^i(\b{k}'',k_0)\phi_b^j(\b{k}+\b{k}'-\b{k}'',k_0')\nn\\
\phantom{=}&\times\int\frac{d^d\b{p}}{(2\pi)^d}
\frac{|\b{p}|^{\epsilon-\delta}|\b{k}-\b{k}''+\b{p}|^{\epsilon-\delta}}
{\left[c^2k_0^2+(\b{k}+\b{p})^2+r\mu^2\right]\left[c^2(k_0')^2+(\b{k}'-\b{p})^2+r\mu^2\right]},
\end{align}
an expression analogous to Eqs.~(\ref{eqA9}-\ref{eqA10}), but with an additional factor $|\b{k}-\b{k}''+\b{p}|^{\epsilon-\delta}$ in the loop integral. Again, the loop integral can be evaluated in the limit of vanishing external momenta. Using Eq.~(\ref{EuclideanIntegral}), we obtain:
\begin{align}
\int\frac{d^d\b{p}}{(2\pi)^d}\frac{|\b{p}|^{2(\epsilon-\delta)}}{(\b{p}^2+r\mu^2)^2}=\frac{2}{(4\pi)^2(2\delta-\epsilon)},
\end{align}
in the limit $\epsilon,\delta\rightarrow 0$, and the corresponding renormalization constant is:
\begin{align}
\delta Z_7^\text{(m)}=-\frac{4v^2\Delta^{-1}}{2\delta-\epsilon}.
\end{align}

\subsection{Long-range correlated disorder strength}

Diagrams on Fig.~\ref{fig:diagrams}(j,l,m) also contribute to the long-range disorder coupling renormalization. Diagram (j) gives:
\begin{align}\label{eqA16}
\delta\Gamma_\text{div}^\text{(j)}=(n+2)\lambda^2 v&\sum_{ab}\int\frac{d^Dk}{(2\pi)^D}\int\frac{d^Dk'}{(2\pi)^D}\int\frac{d^d\b{k}''}{(2\pi)^d}
\,\phi_a^i(-k)\phi_b^j(-k')\phi_a^i(\b{k}'',k_0)\phi_b^j(\b{k}+\b{k}'-\b{k}'',k_0')\nn\\
&\phantom{=}\times|\b{k}-\b{k}''|^{\epsilon-\delta}\int\frac{d^Dp}{(2\pi)^D}\frac{1}
{(c^2p_0^2+\b{p}^2+r\mu^2)\left[c^2p_0^2+(\b{k}-\b{k}''-\b{p})^2+r\mu^2\right]}.
\end{align} 
The interaction term induced by long-range correlated disorder in Eq.~(\ref{Sdis}) can be Fourier transformed to momentum space~\cite{FTLRD}:
\begin{align}\label{FakeFT}
\int d^d\b{x}\,d^d\b{x}'\,d^{\epsilon_\tau}\tau\,d^{\epsilon_\tau}\tau'\, \frac{
\b{\phi}_a^2(\b{x},\tau)\b{\phi}_b^2(\b{x},\tau')}{|\b{x}-\b{x}'|^\alpha}&=
\int\frac{d^Dk}{(2\pi)^D}\int\frac{d^Dk'}{(2\pi)^D}\int\frac{d^d\b{k}''}{(2\pi)^d}
\,\phi_a^i(-k)\phi_b^j(-k')\phi_a^i(\b{k}'',k_0)\nn\\
&\phantom{=}\hspace{10mm}\times\phi_b^j(\b{k}+\b{k}'-\b{k}'',k_0')|\b{k}-\b{k}''|^{\epsilon-\delta},
\end{align}
using $d=4-\epsilon$ and $\alpha=4-\delta$. Comparing with Eq.~(\ref{eqA16}), we see that we can evaluate the loop integral in the limit of zero external momenta:
\begin{align}
\int\frac{d^Dp}{(2\pi)^D}\frac{1}{(c^2p_0^2+\b{p}^2+r\mu^2)^2}=\frac{\Gamma(\epsilon/2)}{(4\pi)^{d/2}}
\int\frac{d^{\epsilon_\tau}p_0}{(2\pi)^{\epsilon_\tau}}\frac{1}{(c^2p_0^2+r\mu^2)^{\epsilon/2}}
=\frac{2}{(4\pi)^2(\epsilon-\epsilon_\tau)},
\end{align}
in the limit $\epsilon,\epsilon_\tau\rightarrow 0$. We correspondingly have:
\begin{align}
\delta Z_8^\text{(j)}=\frac{4(n+2)\lambda^2}{\epsilon-\epsilon_\tau}.
\end{align}

Diagram (l) gives:
\begin{align}
\delta\Gamma_\text{div}^\text{(l)}=-\Delta v&\sum_{ab}\int\frac{d^Dk}{(2\pi)^D}\int\frac{d^Dk'}{(2\pi)^D}\int\frac{d^d\b{k}''}{(2\pi)^d}
\,\phi_a^i(-k)\phi_b^j(-k')\phi_a^i(\b{k}'',k_0)\phi_b^j(\b{k}+\b{k}'-\b{k}'',k_0')\nn\\
&\phantom{=}\times|\b{k}-\b{k}''|^{\epsilon-\delta}\int\frac{d^d\b{p}}{(2\pi)^d}
\frac{1}{\left[c^2(k_0')^2+\b{p}^2+r\mu^2\right]\left[c^2(k_0')^2+(\b{k}-\b{k}''-\b{p})^2+r\mu^2\right]}.
\end{align} 
Once again, the loop integral can performed setting to zero the external momenta:
\begin{align}
\int\frac{d^d\b{p}}{(2\pi)^d}\frac{1}{(\b{p}^2+r\mu^2)^2}=\frac{2}{(4\pi)^2\epsilon},
\end{align}
in the limit $\epsilon\rightarrow 0$, and we obtain:
\begin{align}
\delta Z_8^\text{(l)}=-\frac{4\Delta}{\epsilon}.
\end{align}

Finally, diagram (m) gives a contribution similar to diagram (l), but with an extra $\b{p}$-dependent factor in the loop integrand:
\begin{align}
\delta\Gamma_\text{div}^\text{(m)}=-v^2&\sum_{ab}\int\frac{d^Dk}{(2\pi)^D}\int\frac{d^Dk'}{(2\pi)^D}\int\frac{d^d\b{k}''}{(2\pi)^d}
\,\phi_a^i(-k)\phi_b^j(-k')\phi_a^i(\b{k}'',k_0)\phi_b^j(\b{k}+\b{k}'-\b{k}'',k_0')\nn\\
&\phantom{=}\times|\b{k}-\b{k}''|^{\epsilon-\delta}\int\frac{d^d\b{p}}{(2\pi)^d}
\frac{|\b{k}+\b{k}'-\b{k}''-\b{p}|^{\epsilon-\delta}}{\left[c^2(k_0')^2+\b{p}^2+r\mu^2\right]\left[c^2(k_0')^2+(\b{k}-\b{k}''-\b{p})^2+r\mu^2\right]}.
\end{align} 
In the limit of vanishing external momenta, the loop integral reduces to Eq.~(\ref{DimRegIntegral}), and we have:
\begin{align}
\delta Z_8^\text{(m)}=-\frac{4v}{\delta}.
\end{align}

Finally, diagrams (e) and (f), which contribute to the renormalization of the fermion two-point function and the Yukawa vertex, respectively, are the same as for the clean theory and have been computed before~\cite{Rosenstein1993,Zerf2017}.

\section{Absence of fermionic short-range disordered fixed point at $\mathcal{O}(\sqrt{\epsilon_\tau})$ in the chiral Ising GNY  model}
\label{app:NoPoint}

In the random-mass chiral Ising GNY model ($n=1$), we found a single SDFP at one-loop order [Eq.~(\ref{SDFP_m=0})], by contrast with the chiral XY and Heisenberg models where we found two SDFPs [Eq.~(\ref{SDFP1})]. This is a consequence of the accidental degeneracy of the system of equations $\beta_{\lambda^2}/\lambda^2=0$, $\beta_\Delta/\Delta=0$ in the bosonic limit $h^2=0$. In the bosonic theory, this accidental degeneracy is lifted at two-loop order, which leads to a SDFP with $\lambda_*^2,\Delta_*\sim\c{O}(\sqrt{\epsilon_\tau})$ for a fixed ratio $\epsilon/\epsilon_\tau$~\cite{Boyanovsky1982,boyanovsky1983,Lawrie1984}. Setting $\epsilon/\epsilon_\tau=2$, we investigate the possibility of an additional fermionic SDFP with $\lambda_*^2,\Delta_*\sim\c{O}(\sqrt{\epsilon_\tau})$ in the random-mass chiral Ising GNY model.

At higher loop orders, for a reason that will become clearer towards the end of this Appendix, it is technically more convenient~\cite{Boyanovsky1982,Lawrie1984} to work with rescaled couplings $\widetilde{\lambda}^2$ and $\widetilde{h}^2$, defined via $\lambda^2=c^{\epsilon_\tau}\widetilde{\lambda}^2$ and $h^2=c^{\epsilon_\tau}\widetilde{h}^2$. Using Eqs.~(\ref{betac2}-\ref{betah2formal}), the beta functions for those rescaled couplings are:
\begin{align}
\beta_{\widetilde{\lambda}^2}&=\left(-\epsilon_\tau+2\gamma_4-\gamma_5+\textstyle{\frac{\epsilon_\tau}{2}}(\gamma_3-\gamma_4)\right)\widetilde{\lambda}^2,\label{bl2tilde}\\
\beta_{\widetilde{h}^2}&=\left(-\epsilon_\tau+2(\gamma_2-\gamma_6)+\gamma_4+\textstyle{\frac{\epsilon_\tau}{2}}(\gamma_3-\gamma_4)\right)\widetilde{h}^2.\label{bh2tilde}
\end{align}
At one-loop order, those beta functions reduce to those previously found [Eqs.~(\ref{bl2}-\ref{bh2})] with $\lambda\rightarrow\widetilde{\lambda}$ and $h\rightarrow\widetilde{h}$. Indeed, there is no change in the divergent part of the one-loop effective action in the limit $\epsilon_\tau\rightarrow 0$, and thus in the $\overline{\text{MS}}$ renormalization constants, and the terms $\frac{\epsilon_\tau}{2}(\gamma_3-\gamma_4)$ in Eqs.~(\ref{bl2tilde}-\ref{bh2tilde}) are dropped at this order. At two-loop order, ignoring these latter terms for now, the beta functions for $\widetilde{\lambda}^2$, $\Delta$, and $\widetilde{h}^2$ read:
\begin{align}
\beta_{\widetilde{\lambda}^2}&=-\epsilon_\tau\widetilde{\lambda}^2+6(3\widetilde{\lambda}^2-2\Delta)\widetilde{\lambda}^2+2N\widetilde{h}^2\widetilde{\lambda}^2-N\widetilde{h}^4+(\text{cubic in }\widetilde{h}^2,\widetilde{\lambda}^2,\Delta),
\\
\beta_\Delta&=-2\epsilon_\tau\Delta+4(3\widetilde{\lambda}^2-2\Delta)\Delta+2N\widetilde{h}^2\Delta+(\text{quadratic in }\widetilde{h}^2,\widetilde{\lambda}^2,\Delta)\times\Delta,\label{bD2L}\\
\beta_{\widetilde{h}^2}&=-\epsilon_\tau \widetilde{h}^2+(N+3)\widetilde{h}^4+(\text{quadratic in }\widetilde{h}^2,\widetilde{\lambda}^2,\Delta)\times \widetilde{h}^2.\label{bh22L}
\end{align}
where the form of the two-loop term in Eq.~(\ref{bD2L}) follows from the fact that a disorder vertex cannot be generated perturbatively from a clean theory. Similarly, Eq.~(\ref{bh22L}) follows from the fact that a Yukawa vertex cannot be generated from a theory of decoupled bosons and fermions.

We expand the fixed-point couplings $\widetilde{\lambda}^2_*$, $\Delta_*$, and $\widetilde{h}^2_*$ in increasing powers of $\epsilon_\tau$:
\begin{align}
\widetilde{\lambda}^2_*=\lambda^2_1+\lambda^2_2+\ldots,\hspace{10mm}
\Delta_*=\Delta_1+\Delta_2+\ldots,\hspace{10mm}
\widetilde{h}^2_*=h^2_1+h^2_2+\ldots,
\end{align}
where the leading power for each coupling remains to be determined. The SDFP (\ref{SDFP_m=0}) previously found was obtained assuming that $3\lambda^2_1-2\Delta_1\neq 0$, which gives $\lambda_1^2,\Delta_1,h_1^2\propto\epsilon_\tau$. Here we consider the possibility that $3\lambda^2_1-2\Delta_1=0$, with $\lambda^2_1,\Delta_1\propto\sqrt{\epsilon_\tau}$~\cite{Boyanovsky1982,boyanovsky1983,Lawrie1984}. First, in Eq.~(\ref{bh22L}), the two-loop term is at most $\propto\epsilon_\tau \widetilde{h}^2$, thus the equation $\beta_{\widetilde{h}^2}=0$ may in general be solved to $\mathcal{O}(\epsilon_\tau^2)$ to yield a nontrivial solution $h_1^2\propto\epsilon_\tau\neq 0$. In fact, $\beta_{\widetilde{h}^2}$ also contains the term $\frac{\epsilon_\tau}{2}(\gamma_3-\gamma_4)\widetilde{h}^2$ [see Eq.~(\ref{bh2tilde})], but at leading order this term is $\c{O}(\epsilon_\tau^{5/2})$ and does not affect $h_1^2$.

At leading order, the equations $\beta_{\widetilde{\lambda}^2}=0$ and $\beta_\Delta=0$ become:
\begin{align}
0&=-\epsilon_\tau\lambda^2_1+6(3\lambda_2^2-2\Delta_2)\lambda_1^2+2Nh_1^2\lambda_1^2-Nh_1^4+(\text{cubic in }h^2_1,\lambda^2_1,\Delta_1),\label{eql22L}\\
0&=-2\epsilon_\tau\Delta_1+4(3\lambda_2^2-2\Delta_2)\Delta_1+2Nh_1^2\Delta_1+(\text{quadratic in }h^2_1,\lambda^2_1,\Delta_1)\times\Delta_1.
\end{align}
These equations may in general be solved to $\c{O}(\epsilon_\tau^{3/2})$ to yield nontrivial solutions $\lambda_1^2,\Delta_1\propto\sqrt{\epsilon_\tau}$, with $\lambda_2^2,\Delta_2\propto\epsilon_\tau$. As with $\beta_{\widetilde{h}^2}$, Eq.~(\ref{eql22L}) in fact contains the additional term $\frac{\epsilon_\tau}{2}(\gamma_3-\gamma_4)\widetilde{\lambda}^2$ on the right-hand side [see Eq.~(\ref{bl2tilde})], but at leading order this term is $\c{O}(\epsilon_\tau^2)$ and does not affect $\lambda_1^2,\Delta_1$.

Thus far we have seen that a common zero of $\beta_{\widetilde{\lambda}^2},\beta_{\widetilde{h}^2},\beta_\Delta$ with $\widetilde{\lambda}^2_*,\Delta_*\sim\c{O}(\sqrt{\epsilon_\tau})$ and $\widetilde{h}_*^2\sim\c{O}(\epsilon_\tau)$ is in principle possible at two-loop order. We now turn to the remaining equation, $\beta_{c^2}=0$. At two-loop order, the beta function for $c^2$ reads:
\begin{align}\label{bc2_app}
\beta_{c^2}=-2\Delta c^2+\widetilde{h}^2\left[N(c^2-1)+c^2(2f(c^2)-1)\right]+\beta_{c^2}^\text{(2L)},
\end{align}
where the two-loop part,
\begin{align}
\beta_{c^2}^\text{(2L)}=\left(2\gamma_1^\text{(2L)}-2\gamma_2^\text{(2L)}-\gamma_3^\text{(2L)}+\gamma_4^\text{(2L)}\right)c^2,
\end{align}
depends on $\gamma_i^\text{(2L)}$, $i=1,\ldots,4$, the two-loop contributions to the anomalous dimensions $\gamma_i=d\ln Z_i/d\ln \mu$. These contributions are quadratic in the couplings $\widetilde{h}^2,\widetilde{\lambda}^2,\Delta$, but may have a nontrivial dependence on $c^2$. We separate $\beta_{c^2}^\text{(2L)}$ into a purely bosonic part and a part depending on the Yukawa coupling:
\begin{align}\label{bc2_app_2}
\beta_{c^2}^\text{(2L)}=(\text{quadratic in }\widetilde{\lambda}^2,\Delta)\times f_1(c^2)c^2
+(\text{linear in }\widetilde{h}^2,\widetilde{\lambda}^2,\Delta)\times \widetilde{h}^2f_2(c^2)c^2,
\end{align}
where $f_1$ and $f_2$ are potentially nontrivial functions of $c^2$. We look for solutions $c_*^2$ to the equation $\beta_{c^2}=0$, evaluated at $\widetilde{\lambda}^2_*,\Delta_*\sim\c{O}(\sqrt{\epsilon_\tau})$ and $\widetilde{h}_*^2\sim\c{O}(\epsilon_\tau)$. Since $c^2$ is not a perturbative coupling, we assume $c_*^2\sim\mathcal{O}(1)$, as in the DFPs studied in the rest of the paper. The first term in (\ref{bc2_app}) is then $\mathcal{O}(\sqrt{\epsilon_\tau})$ while the remaining terms are $\mathcal{O}(\epsilon_\tau)$, so there is no consistent $c_*^2\neq 0$ solution.

We can at last look for a fixed point with $c_*^2=0$, such that $\beta_{c^2}=-N\widetilde{h}_*^2+\beta_{c^2}^\text{(2L)}$. If we can show that $f_1(c^2)=\text{const.}$, this solution is again inconsistent at leading order in $\epsilon_\tau$ since, even if $f_2(c^2)c^2$ remains finite in the limit $c_*^2\rightarrow 0$, the $\widetilde{h}^2$-dependent term in Eq.~(\ref{bc2_app_2}) is then $\mathcal{O}(\epsilon_\tau^{3/2})$. We now proceed to show that $f_1(c^2)$ is in fact independent of $c^2$. To do so, we can restrict ourselves to the purely bosonic theory with $\widetilde{h}^2=0$. In this case one has $\gamma_1=\gamma_2=0$ at all loop orders, since the fermions decouple and remain free fields, and we need only consider the contributions of $\gamma_3^\text{(2L)}$ and $\gamma_4^\text{(2L)}$ to $\beta_{c^2}^\text{(2L)}$, i.e., two-loop corrections to the boson two-point function in the bosonic theory. These are essentially the standard double tadpole and sunset diagrams of two-loop $\phi^4$ theory, but with $V_\lambda$ self-interaction vertices and $V_\Delta$ disorder vertices such that $V_\lambda+V_\Delta=2$. Schematically, these corrections are of the form:
\begin{align}\label{2loop}
\delta D(k_0,\b{k})\propto\left(c^{\epsilon_\tau}\widetilde{\lambda}^2\right)^{V_\lambda}\Delta^{V_\Delta}
\int d^{\epsilon_\tau}p_0\,d^{\epsilon_\tau}p_0'\,d^d\b{p}\,d^d\b{p}'\left[\delta^{(\epsilon_\tau)}(k_0,p_0,p_0')\right]^{V_\Delta}I(ck_0,\b{k};cp_0,cp_0',\b{p},\b{p}'),
\end{align}
where $k=(k_0,\b{k})$ is the external momentum, $p=(p_0,\b{p})$ and $p'=(p_0',\b{p}')$ are the two independent loop momenta, and $\left[\delta^{(\epsilon_\tau)}(k_0,p_0,p_0')\right]^{V_\Delta}$ symbolizes the fact that each disorder vertex is accompanied by an $\epsilon_\tau$-dimensional delta function involving linear combinations of the frequencies $k_0,p_0,p_0'$ in the diagram (see Fig.~\ref{fig:feynrules}). Performing the change of integration variables $p_0\rightarrow\widetilde{p}_0=cp_0$, $p_0'\rightarrow\widetilde{p}_0'=cp_0'$, and using the property $\delta^{(\epsilon_\tau)}(q_0/c)=c^{\epsilon_\tau}\delta^{(\epsilon_\tau)}(q_0)$, Eq.~(\ref{2loop}) becomes:
\begin{align}
\delta D(k_0,\b{k})&\propto c^{(V_\lambda+V_\Delta-2)\epsilon_\tau}
\left(\widetilde{\lambda}^2\right)^{V_\lambda}\Delta^{V_\Delta}\nn\\
&\hspace{15mm}\times\int d^{\epsilon_\tau}\widetilde{p}_0\,d^{\epsilon_\tau}\widetilde{p}_0'\,d^d\b{p}\,d^d\b{p}'\left[\delta^{(\epsilon_\tau)}(ck_0,\widetilde{p}_0,\widetilde{p}_0')\right]^{V_\Delta}I(ck_0,\b{k};\widetilde{p}_0,\widetilde{p}_0',\b{p},\b{p}'),
\end{align}
which, since $V_\lambda+V_\Delta=2$, depends on $c$ only through $c^2k_0^2$. Since the latter appears in the unperturbed propagator (\ref{BosPropag}), $\gamma_3^\text{(2L)}$ and $\gamma_4^\text{(2L)}$, and thus $f_1(c^2)$ in Eq.~(\ref{bc2_app_2}), are necessarily independent of $c^2$. Similar reasoning shows that counter-term insertions in one-loop diagrams do not generate a dependence on $c^2$ either. According to the argument above, a fixed point with $c_*^2=0$ is thus impossible.

\section{Log-periodic scaling laws from limit-cycle criticality}
\label{app:LimitCycleScaling}

In this last Appendix we derive the effects of limit-cycle criticality on scaling laws. We focus on the uniform static susceptibility $\chi$, but the derivation can be extended to other thermodynamic observables. Ignoring corrections to the dynamic critical exponent, the two-point function of the order parameter $\chi(\b{q})=\langle\b{\phi}(\b{q})\cdot\b{\phi}(-\b{q})\rangle$ obeys the scaling relation $\chi(\b{q},r(0))=e^{(2-\eta_\phi)\ell}\chi(e^\ell\b{q},r(\ell))$. Here we switched from the RG scale $\mu$ to the infrared scale parameter $\ell\sim-\ln\mu$;  $r(0)$ and $r(\ell)$ are the bare and renormalized tuning parameters for the transition, respectively. We are also using the fact that at one-loop order, $\eta_\phi$ depends only on $h^2_*$, and is thus constant everywhere on the limit cycle. The tuning parameter $r(\ell)$ is renormalized according to the equation:
\begin{align}\label{drdl}
\frac{dr(\ell)}{d\ell}=[2-\gamma_{4}(\b{g}(\ell))+\gamma_r(\b{g}(\ell))]r(\ell)=[2-\gamma_{m^2}(\b{g}(\ell))]r(\ell).
\end{align}
In turn, $\b{g}(\ell)=\bigl(c^2,h^2,\lambda^2,\Delta,v)$, a vector of renormalized couplings, flows according to the obtained (infrared) beta-functions:
\begin{align}\label{dgdl}
\frac{d\b{g}(\ell)}{d\ell}=\boldsymbol{\beta}(\b{g}(\ell)).
\end{align}
Lets denote $r(0)$ by $r$, and define $\ell_r$ such that $r(\ell_r)=r_0$. Then choosing $\ell=\ell_r$, the uniform thermodynamic susceptibility is $\chi(\b{q}=0,r)\sim e^{(2-\eta_\phi)\ell_r}$. The goal is to determine $\ell_r$ as a function of $r$. From Eq.~(\ref{drdl}), we find
\begin{align}\label{logr0r}
\ln\left(\frac{r_0}{r}\right)=\int_0^{\ell_r}d\ell\,[2-\gamma_{m^2}(\b{g}(\ell))].
\end{align}
For initial values of couplings $\b{g}(0)$ such that they are on the limit cycle, the integration of Eq.~(\ref{dgdl}) gives periodic functions $\b{g}(l)$ with period $\ell_\text{LC}$, which can then be expanded as a Fourier series:
\begin{align}
\b{g}(\ell)=\sum_{n=-\infty}^\infty\b{g}_n\,e^{2\pi in\ell/\ell_\text{LC}},
\end{align}
with $\b{g}_n=\b{g}_{-n}^*$ since $\b{g}(\ell)$ is real. At one-loop order, $\gamma_{m^2}$ is linear in the couplings, $\gamma_{m^2}(\b{g}(\ell))=\b{a}\cdot\b{g}(\ell)$. Performing the integration over $\ell$ in Eq.~(\ref{logr0r}), we obtain:
\begin{align}\label{lreqn}
\ln\left(\frac{r_0}{r}\right)=\nu_\text{LC}^{-1}\ell_r-\c{F}(\ell_r),
\end{align}
where
\begin{align}\label{nuLCinv}
\nu_\text{LC}^{-1}=2-\b{a}\cdot\langle\b{g}\rangle_\text{LC},
\end{align}
is an effective inverse correlation-length exponent associated with the critical limit cycle, and the function $\c{F}$ defined as
\begin{align}\label{FLC}
\c{F}(\ell_r)=\b{a}\cdot\int_0^{\ell_r}d\ell\left[\b{g}(\ell)-\langle\b{g}\rangle_\text{LC}\right]
=\sum_{n\neq 0}\b{a}\cdot\b{g}_n\frac{e^{2\pi in\ell/\ell_\text{LC}}-1}{2\pi in/\ell_\text{LC}},
\end{align}
is periodic in $\ell_r$ with the period $\ell_\text{LC}$ of the limit cycle. In Eqs.~(\ref{nuLCinv}-\ref{FLC}), $\langle\b{g}\rangle_\text{LC}$ is the ``center'' of the limit cycle, i.e., the average of $\b{g}(\ell)$ over one period,
\begin{align}
\langle\b{g}\rangle_\text{LC}=\frac{1}{\ell_\text{LC}}\int_0^{\ell_\text{LC}}d\ell\,\b{g}(\ell),
\end{align}
and coincides with the zeroth Fourier component $\b{g}_0$. For limit cycles with inversion symmetry with respect to the enclosed unstable-focus fixed point $\b{g}_*$ (see Sec.~\ref{sec:bifurc2}), $\nu_\text{LC}$ would coincide with the correlation-length exponent at this fixed point.

If the limit cycle is small, e.g., near the Hopf bifurcation, we see from Eq.~(\ref{FLC}) that $\c{F}$ is also small, in which case Eq.~(\ref{lreqn}) can be solved perturbatively in the radius of the limit cycle. To first order in this radius, we thus obtain:
\begin{align}
\ell_r\approx\nu_\text{LC}\ln\left(\frac{r_0}{r}\right)+\nu_\text{LC}\c{F}\left(\nu_\text{LC}\ln\left(\frac{r_0}{r}\right)\right).
\end{align}
Substituting into $\chi\equiv\chi(\b{q}=0,r)\sim e^{(2-\eta_\phi)\ell_r}$, and consistently working to first order in $\c{F}$, we obtain:
\begin{align}
\chi\sim|r|^{-\gamma_\text{LC}}\left[1+\gamma_\text{LC}\c{F}\left(\nu_\text{LC}\ln\left(\frac{r_0}{r}\right)\right)\right],
\end{align}
which is Eq.~(\ref{ChiLimitCycle}) in the main text, where we have defined $\gamma_\text{LC}=(2-\eta_\phi)\nu_\text{LC}$.

\numberwithin{equation}{section}
\numberwithin{figure}{section}

\bibliography{DisorderedGNYpaper}

\end{document}